# A novel S-shape based NURBS interpolation with acc-jerk-Continuity and round-off error elimination

Yifei Hu[1,2,3], Xin Jiang[1,2,4,*], Guanying Huo[1,2,*], Cheng Su[1,2,3], Bolun Wang[1,2,3], Hexiong Li[1,2], Zhiming Zheng[1,2]

[1]Key Laboratory of Mathematics, Informatics and Behavioral Semantics (LMIB), School of Mathematics Science, Beihang University, Beijing, China

[2]Peng Cheng Laboratory, Shenzhen, Guangdong, China

[3]Shenyuan honor college, Beihang University, Beijing, China

[4]Beijing Advanced Innovation Center for Big Data and Brain Computing (BDBC), Beihang University, Beijing, China

Corresponding author: Xin Jiang (e-mail: jiangxin@buaa.edu.cn ). Guanying Huo (e-mail: gyhuo@buaa.edu.cn )

## Abstract

Feedrate scheduling is a key step in computer numerical control (CNC) machining, as it has a close relationship with machining time and surface quality, and has now become a hot issue in industry and academia. To reduce high chord errors and round-off errors, and generate continuous velocity, acceleration, and jerk profile of parametric interpolation, a novel and complete S-shape based feedrate scheduling algorithm is presented in this paper. The algorithm consists of three modules: bidirectional scanning module, velocity scheduling module and round-off error elimination module. The bidirectional scanning module with the limitations of chord error, normal acceleration/jerk and command feedrate aims to guarantee the continuity of the feed rate at the junctions between successive NURBS sub-curves. After the NURBS sub-curves have been classified into two cases, the velocity scheduling module firstly calculates the actual maximum federate, and then generates the feed rate profiles of all NURBS sub-curves according to our velocity scheduling function. Later, the round-off error elimination module is proposed to make the total interpolating time become an integer multiple of the interpolation period, which leads to the elimination of round-off errors. Finally, benchmarks are conducted to verify the applicability of the proposed method compared with some other methods.

## Introduction

In multi-axis computer numerical control (CNC) machining, most of the tool paths are defined by $G01$ blocks [1]. However, due to the following two reasons, the $G01$ blocks cannot be used in machining processes directly. Firstly, to completely stop at the corners and change the feed direction, frequent acceleration and deceleration processes are inevitable [2]. Secondly, the derivates are not the same at the junctions of line and circular segments, and the feed changes abruptly when the tool moves across the junction [3]. Both reasons can have a serious impact on machining quality. Another problem is that the interpolating time cannot be guaranteed to be an integer multiple of the interpolation period, which may introduce round-off errors. Since the existence of round-off errors will affect the machining accuracy and motion smoothness, how to compensate or eliminate round-off errors should not be ignored in feedrate scheduling.

To solve the first problem, there are two main types of methods. For the first type of methods, several transition schemes [3-8] are proposed to smooth the toolpath firstly, then feedrate scheduling is applied to transition curves, linear and circular segments. However, in this type of method, a local curve is utilized to connect two short line segments or circular segments. Since the NC programs consist of a large number of short line segments or circular arc segments, a lot of computation and memory are needed. For the second type of method, a global curve is used to approximate all linear and circular segments, and then the feedrate is planned for this global curve. Since this method has reduced the amount of geometric information transferred from the CAD system to the CNC system and the errors between CAD/CAM and CNC, it became the main research direction.

Shpitalni et al. [9] first proposed parametric curve interpolation. Then, several interpolation methods were proposed to further increase the machining accuracy, such as method based on Taylor's expansion [10], predictor-corrector interpolator(PCI) algorithm [11], quintic spline interpolation method [12], and inverse length functions (ILF) method [13]. Nevertheless, the feedrate is constant in the above methods, which leads to massive chord error and acceleration/deceleration(ACC/DEC). Later, in [14], an adaptive-feed rate interpolator whose chord error does not exceed a given threshold was proposed. Jeon et al. [15] proposed a generalized approach for generating velocity profiles with the desired characteristics of acceleration and deceleration. In [16], a curvature-based interpolation algorithm was presented to generate the feedrate profile according to the curvature of curves.

However, if the curvature of a curve changes abruptly, the jerk may exceed the limit capacity of the machine, causing the vibration of the machine, which will reduce the machining accuracy. To cope with this problem, several adaptive interpolation methods considering the jerk limit have been proposed in [17-27]. Erkorkmaz et al. [17] first introduced an S-shape federate scheming to generate jerk limited trajectory. In [24], an optimized S-shaped $C^2$ quintic feedrate planning scheme was proposed to generate an acc-jerk-limited feedrate profile. Xu et al. [25] classified the NURBS blocks into three types firstly, then generated federate profiles with a confined jerk, acceleration, and command feed rate for each block. In [26], a novel S-curve acceleration and deceleration (ACC/DEC) control model was proposed, and the discretization of the Acc/Dec process was investigated. Jia et al. [27] introduced the concept of feedrate-sensitive regions firstly. To reduce the frequency of feedrate changes, the feedrate kept constant at feedrate-sensitive regions and varied smoothly within parts of the transition regions, which are beneficial to the machining quality.

Although the jerk of these methods will not exceed the limit, the machining accuracy will be affected due to the drastic change of jerk. To further generate continuous and smooth acceleration transition profiles, two acceleration smoothing algorithms based on the jounce limit were proposed in [28, 29]. Since the calculation process is complicated, it is not suitable for parametric curve interpolation. In [30], a sine-curve velocity profile was proposed to generate the federate profile, which constructed a jerk-limited and continuous feed rate profile. Wang et al. [31] presented a trigonometric velocity scheduling algorithm based on two-time look-ahead interpolation. In [32], a novel approach for non-uniform rational B-spline (NURBS) interpolation through the integration of an acc-jerk-continuous-based control method and a look-ahead algorithm was proposed. However, most methods are based on the trigonometric profile, and the machining process becomes less efficient as only one point can reach the given maximum value of the acceleration or jerk [33].

To cope with the second problem. Xu et al. [25] proposed a variable-jerk compensation strategy to compensate for the round-off error, which makes the feed rate more continuous, and the arc increment more precise. In [34-35], similar time rounding and error compensation methods were proposed, which lead to the discontinuity of acceleration profiles. In [36], an optimized feedrate scheduling consisting of initial feed rate scheduling and parameters calculation of round-off error compensation based on the S-shaped ACC/DEC algorithm was introduced. Although the interpolating time can become an integer times of the interpolation period, some kinematic characteristics might exceed the limit. Ni et al. [37] proposed a time-rounding-up schemes to make the total interpolation time of each curve segment an integer multiple of interpolation period, in which four feedrate scheduling methods are designed for different situations. Although the round-off error can be eliminated for most situations, when these methods are not applicable the method in [36] will be adopted, which will cause the same problem in [36].

Considering chord error, round-off error and kinematic characteristics, a novel S-shape velocity scheduling function based on trigonometric function is proposed in this paper. Due to the advantages of the NURBS curve in complex surface modeling, the default parametric curve in this paper is NURBS type [38]. In fact, the feedrate scheduling framework is also suitable for other parametric curve forms. Our algorithm consists of three modules: bidirectional scanning module, velocity scheduling module and round-off error elimination module. For short NURBS blocks, the bidirectional scanning module generates the actual maximum velocity by solving up to two univariate cubic equations, then adjusting the end velocity accordingly if necessary. After the bidirectional scanning module, the NURBS blocks may be two situations of the path: the case of acceleration/deceleration (ACC-or-DEC case) and the case of acceleration–uniform–deceleration or acceleration–deceleration (ACC-and-DEC case). For the former, since the actual maximum velocity is known, the relevant time parameters of the velocity profile can be easily obtained; for the latter, the actual maximum velocity can be obtained by solving up to one univariate quadratic equation and two univariate quartic equations, and then the relevant time parameters of the velocity profile can also be obtained. In round-off error elimination module, the actual maximum velocity of a NURBS block of ACC-and-DEC case is reduced appropriately so that the total interpolating time becomes an integral multiple of the interpolation period without changing the initial and terminal velocity of this NURBS block.

The rest of this paper is as follows: Section 2 introduces the basic knowledge of NURBS and the process of generating NURBS blocks. Then, an improved S-shape velocity scheduling function based on trigonometric function, the details of the bidirectional scanning module, velocity scheduling module, and round-off error elimination module are introduced in Section 3. In Section 4, the proposed method is verified in simulations, and several classical methods are compared with our algorithm in different ways. Finally, the conclusions are given in Section 5.

## 2. Preliminary knowledge

In this section, the definition of NURBS curve is first reviewed. Then the process of generating NURBS sub-curves is given.

### 2.1 The definition of NURBS curve

A $p^{th}$-order NURBS curve can be expressed as follows [37]:

$$C(u)=\frac{A(u)}{B(u)}=\frac{\sum_{i=0}^{n}N_{i,p}(u)w_iP_i}{\sum_{i=0}^{n}N_{i,p}(u)w_iP_i} \tag{1}$$

Where $\{P_i\}$ are the control points, $\{w_i\}$ are the weights of $\{P_i\}$, $n+1$ is the number of control points and $\{N_{i.p}(u)\}$ are the $p^{th}$-order B-spline basis functions defined on the non-periodic and heterogeneous knot vector $U=\{u_0,u_1,...,u_{n+p+1}\}$. Then $\{N_{i.p}(u)\}$ is recursively defined as follows:

$$N_{i,0}(u)=\begin{cases}1 & if\ u_i\le u\le u_{i+1}\\ 0 & otherwise\end{cases} \tag{2}$$

$$\begin{aligned}N_{i,p}(u)&=\frac{u-u_i}{u_{i+p}-u_i}N_{i,p-1}(u)+\frac{u_{i+p+1}-u}{u_{i+p+1}-u_{i+1}}N_{i+1,p-1}(u)\\ i&=0,1,...,n.\end{aligned} \tag{3}$$

Usually, the range of values allowed for the parameter $u$ are $u_0=0$ and $u_{n+p+1}$=1, and $w_i>0$ for all $i$. Since the first and second derivatives of a NURBS curve are involved in subsequent calculations, their calculation formula are given as follows:

$$C'(u) = \frac{A'(u) - B'(u)C(u)}{B(u)} \tag{4}$$

$$C''(u) = \frac{A''(u) - 2C'(u)B'(u) - C(u)B''(u)}{B(u)} \tag{5}$$

2.2 The division of NRUBS curve

Before feed rate scheduling, the NURBS curve should be divided into several sub-curves. In this paper, we divide the NURBS curve according to the maximum allowable feed rate directly. For each parameter $u_i$ of a NURBS curve, the corresponding maximum allowable feed rate can be obtained considering the chord error, centripetal acceleration and jerk limitations. Then, the parameters whose corresponding maximum allowable feed rate are minimum are used as the breakpoints to divide the NURBS curve into several sub-curves.

The method in [30] for calculating the corresponding maximum allowable feed rate is adopted in this paper. The calculation formulas are as follows:

$$v_{i,1} = \frac{2}{T_s}\sqrt{2\rho_i\delta - \delta^2} \tag{6}$$

$$v_{i,2} = \sqrt{A_n\rho_i} \tag{7}$$

$$v_{i,3} = \sqrt[3]{J_n^2\rho_i} \tag{8}$$

$$v_{i,\ \mathrm{limit}} = \min(F, v_{i,1}, v_{i,2}, v_{i,3}) \tag{9}$$

Where $T_s$ is the interpolation period, $\rho_i$ is the curvature radius of parameter $u_i$, $\delta$

is the maximum chord error, $A_n$, $J_n$ and $F$ are the centripetal acceleration, centripetal jerk and feed rate limitations respectively.

After these breakpoints are detected, the NURBS curve is divided into several sub-curves. To know the arc length of each sub-curve, Simpson's method is proposed. Finally, for each sub-curve $C_i$, the corresponding features vector $(u_{\mathrm{strar}}, u_{end}, v_{start}, v_{end}, l_i)$ are obtained. Where $u_{start}$ and $u_{end}$ are the parameters corresponding to the start and end points of this sub-curve, $v_{start}$ and $v_{end}$ are the corresponding maximum allowable feed rate and $l_i$ is the arc length.

## 3. Proposed method

In this section, the proposed method is detailed. Firstly, the improved S-shape velocity scheduling function is introduced, where the arc length of each sub-curve $l_{\mathrm{i}}$ is supposed to be long enough, so the feed rate can accelerates from the start velocity $v_{start}$ to the command velocity $F$, maintain the command velocity $F$ for a period of time and finally decelerates to the end velocity $v_{end}$. However, in practice, if the arc length is not long enough, the feed rate may not reach the command velocity before it needs to decelerate. Another condition is that when the arc length is too short, even the feed rate keep accelerating or decelerating, it cannot reach the end velocity before reaching the endpoint of the sub-curve. To solve these problems, the bidirectional scanning module and velocity scheduling module are applied in our algorithm. In bidirectional scanning module, the start velocity and end velocity of sub-curves will be adjusted based on the improved S-shape velocity scheduling function if necessary. In velocity scheduling module, the actual maximum velocity of sub-curves is calculated firstly, then the corresponding jerk, acceleration, velocity and displacement of any time $t$ of each sub-curve can be obtained easily. Since the interpolating time is not an integer multiple of the interpolation period in most cases, round-off error will be introduced. In our algorithm, a round-off error elimination module is proposed, by reducing the actual maximum velocity of a certain sub-curve, the total interpolating time becomes an integer multiple of the interpolation period. Meanwhile, the interpolating time is extended by no more than one interpolation period.

### 3.1 Improved S-shape velocity scheduling function

In [32], a five phases S-shape velocity scheduling function based on trigonometric function is applied. It has been proved that when the jerk profile is continuous, the chord errors can be reduced and the vibration of the machine tool can be suppressed to a certain extent. However, due to the fact that the acceleration can't maintain at the maximum for a period of time, the velocity can't be changed fast. For this reason, an improved seven phases S-shape velocity scheduling function is introduced, which is shown as Fig. 1. Similar to [32], the arc length $L$ is assumed to be long enough, then the feed rate can accelerates from the start velocity $v_s$ to the command velocity $F$, maintain the velocity $F$ is for a period of time, and finally decelerates to the end velocity $v_e$.

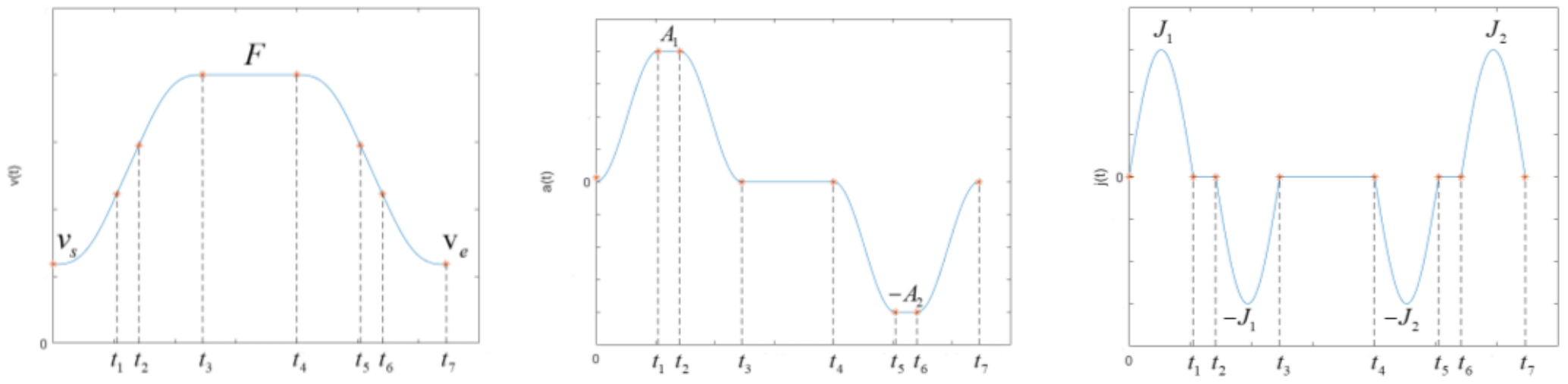

**Fig.1** The feed rate, acceleration and jerk profile.

It is clear that the proposed velocity profile is more continuous compared with traditional S-shaped velocity profile, and the acceleration can maintain at the maximum for a while. Its acceleration equation can be given as formula (10):

$$A(t)=\begin{cases} A_1(1-\cos\dfrac{\pi t}{T_1})/2 & 0\le t<t_1 \\ A_1 & t_1\le t<t_2 \\ A_1(1+\cos\dfrac{\pi(t-t_2)}{T_3})/2 & t_2\le t<t_3 \\ 0 & t_3\le t<t_4 \\ -A_2(1-\cos\dfrac{\pi(t-t_4)}{T_5})/2 & t_4\le t<t_5 \\ -A_2 & t_5\le t<t_6 \\ -A_2(1+\cos\dfrac{\pi(t-t_6)}{T_7})/2 & t_6\le t<t_7 \end{cases} \tag{10}$$

Where $T_1=T_3$ , $T_5=T_7$ and $t_i=\sum_{j=1}^{i}T_j$ , $A_1$ and $A_2$ are the actual maximum acceleration in accelerating period and decelerating period respectively. Differentiating equation (10), the jerk equation can be obtained as formula (11):

$$
J(t)=\begin{cases}
\dfrac{A_1\pi}{2T_1}\sin\dfrac{\pi t}{T_1} & 0\le t<t_1\\
0 & t_1\le t<t_2\\
-\dfrac{A_1\pi}{2T_3}\sin\dfrac{\pi(t-t_2)}{T_3} & t_2\le t<t_3\\
0 & t_3\le t<t_4\\
-\dfrac{A_2\pi}{2T_5}\sin\dfrac{\pi(t-t_4)}{T_5} & t_4\le t<t_5\\
0 & t_5\le t<t_6\\
\dfrac{A_2\pi}{2T_7}\sin\dfrac{\pi(t-t_6)}{T_7} & t_6\le t<t_7
\end{cases}
\tag{11}
$$

Integrating equation (10) gives the velocity equation as formula (12):

$$
v(t)=\begin{cases}
v_s+\dfrac{A_1}{2}t-\dfrac{A_1T_1}{2\pi}\sin\dfrac{\pi t}{T_1} & 0\le t<t_1\\
v_s+\dfrac{A_1}{2}T_1+A_1(t-t_1) & t_1\le t<t_2\\
v_s+\dfrac{A_1}{2}T_1+A_1T_2+\dfrac{A_1}{2}(t-t_2)+\dfrac{A_1T_3}{2\pi}\sin\dfrac{\pi(t-t_2)}{T_3} & t_2\le t<t_3\\
F & t_3\le t<t_4\\
F-\dfrac{A_2}{2}T_5+\dfrac{A_2T_5}{2\pi}\sin\dfrac{\pi(t-t_4)}{T_5} & t_4\le t<t_5\\
F-\dfrac{A_2}{2}T_5-A_2(t-t_5) & t_5\le t<t_6\\
F-\dfrac{A_2}{2}T_5-A_2T_6-\dfrac{A_2}{2}(t-t_6)-\dfrac{A_2T_7}{2\pi}\sin\dfrac{\pi(t-t_6)}{T_7} & t_6\le t<t_7
\end{cases}
\tag{12}
$$

Take the accelerating process as an example, according to equation (12), one equation can be obtained as:

$$A_1(T_1+T_2)=F-v_s \tag{13}$$

This formula can be written as:

$$A_1=\frac{F-v_s}{T_1+T_2} \tag{14}$$

According to equation (11), the actually maximum jerk in accelerating process is:

$$J_1 = \frac{A_1 \pi}{2T_1} \tag{15}$$

When equation (14) is substituted into equation (15), the actual maximum jerk can be expressed as:

$$J_1 = \frac{(F - v_s)\pi}{2T_1(T_1 + T_2)} \tag{16}$$

Since the actual maximum acceleration and jerk can't excess the tangential acceleration limitation $A_t$ and the tangential jerk limitation $J_t$ respectively, the following inequality should be satisfied.

$$\frac{F - v_s}{T_1 + T_2} \leq A_t \tag{17}$$

$$\frac{(F - v_s)\pi}{2T_1(T_1 + T_2)} \leq J_t \tag{18}$$

To maintain high efficiency, the accelerating period $T_1 + T_2 + T_3$ should be minimized. Then the selection of these values $T_1$, $T_2$ and $T_3$ can be transformed into a simple optimization problem as follows:

$$\min T_1 + (T_1 + T_2)$$
$$s.t. \begin{cases} T_1 + T_2 \geq \dfrac{F - v_s}{A_t} \\ T_1(T_1 + T_2) \geq \dfrac{(F - v_s)\pi}{2J_t} \end{cases} \tag{19}$$

This problem can be solved with an illustration. As shown in Fig.2, the X-axis represents period $T_1$ and the Y-axis represents period $T_1 + T_2$ . Since $T_2 > 0$ , then $T_1 + T_2 > T_1$, which means the optimal point $Q = (x, y)$ must above the line $l_1$. The first constraint in equation (19) can be represented by a parabola $c_1$, then point $Q$ must above the parabola $c_1$. The second constraint in equation (19) can be represented by a line parallel to X-axis, which is noted by $l_2$. Marking the intersection point of $l_1$ and $c_1$ as point $B$. If $l_2$ below the point $B$, the optimal point $Q$ coincides with point $B$; Otherwise, when $l_2$ above the point $B$, the optimal point $Q$ coincides with point $A$, which is the intersection point of $l_2$ and $c_1$. The optimal solution can be represented by formula below:

$$\begin{cases} T_1 = \sqrt{\dfrac{(F - v_s)\pi}{2J_t}}, T_2 = 0 & \text{if } F - v_s \le \dfrac{A_t^2}{2J_t} \\ T_1 = \dfrac{\pi A_t}{2J_t}, T_2 = \dfrac{F - v_s}{A_t} - T_1 & \text{else} \end{cases} \tag{20}$$

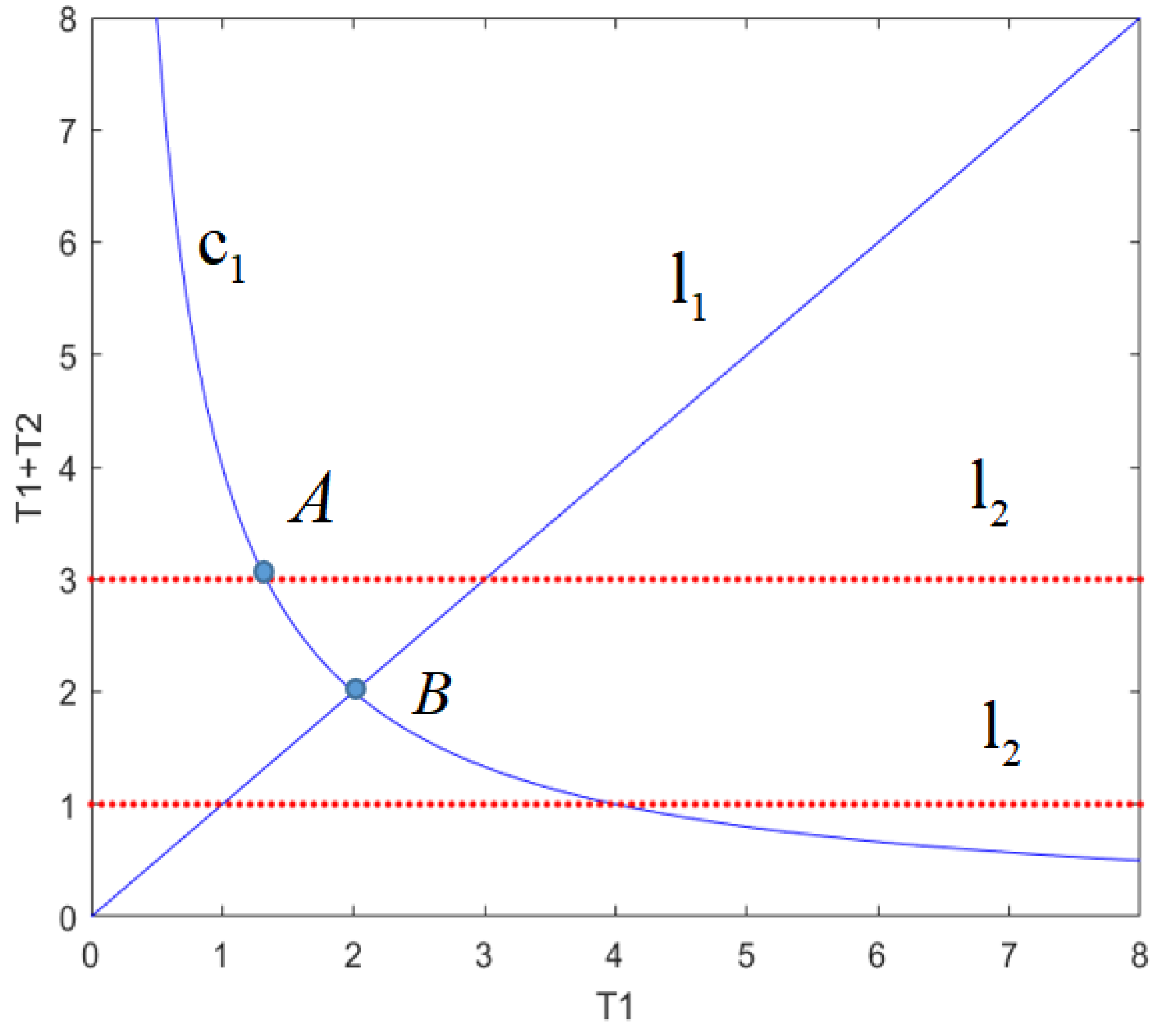

**Fig.2** The schematic diagram of linear programming

To calculate the displacement of accelerating process, integrating Eq. (12) gives the displacement equation as formula (21):

$$S(t)=\begin{cases} v_s t+\frac{A_1}{4}t^2+\frac{A_1T_1^2}{2\pi^2}(\cos\frac{\pi t}{T_1}-1) & 0\le t<t_1 \\ s_1+(v_s+\frac{A_1}{2}T_1-A_1t_1)(t-t_1)+\frac{A_1}{2}(t^2-t_1^2) & t_1\le t<t_2 \\ s_2+(v_s+\frac{A_1}{2}T_1+A_1T_2-\frac{A_1}{2}t_2)(t-t_2)+\frac{A_1}{4}(t^2-t_2^2)-\frac{A_1T_3^2}{2\pi^2}(\cos\frac{\pi(t-t_2)}{T_3}-1) & t_2\le t<t_3 \\ s_3+F(t-t_3) & t_3\le t<t_4 \\ s_4+(F+\frac{A_2}{2}t_4)(t-t_4)-\frac{A_2}{4}(t^2-t_4^2)-\frac{A_2T_5^2}{2\pi^2}(\cos\frac{\pi(t-t_4)}{T_5}-1) & t_4\le t<t_5 \\ s_5+(F-\frac{A_2}{2}T_5+A_2t_5)(t-t_5)-\frac{A_2}{2}(t^2-t_5^2) & t_5\le t<t_6 \\ s_6+(F-\frac{A_2}{2}T_5-A_2T_6+\frac{A_2}{2}t_6)(t-t_6)-\frac{A_2}{4}(t^2-t_6^2)+\frac{A_2T_7^2}{2\pi^2}(\cos\frac{\pi(t-t_6)}{T_7}-1) & t_6\le t<t_7 \end{cases} \quad (21)$$

Where $s_1=v_sT_1+(\frac{1}{4}-\frac{1}{\pi^2})A_1T_1^2$, $s_2=s_1+(v_s+\frac{A_1}{2}T_1)T_2+\frac{A_1}{2}T_2^2$, $s_3=\frac{v_s+F}{2}(2T_1+T_2)$, $s_4=s_3+FT_4$, $s_5=s_4+FT_5+(\frac{1}{\pi^2}-\frac{1}{4})A_2T_5^2$, $s_6=s_5+FT_6-\frac{A_2}{2}T_5T_6-\frac{A_2}{2}T_6^2$ and $s_7=s_4+\frac{v_e+F}{2}(2T_5+T_7)$. According to equation (21), the displacement of accelerating process can be given as:

$$S_1=\frac{(F+v_s)(2T_1+T_2)}{2} \quad (22)$$

Similarly, the period of decelerating process $T_5$, $T_6$ and $T_7$, and the displacement of decelerating process $S_2$ can be obtained. Then the period of constant velocity process can be given as:

$$T_4=\frac{L-S_1-S_2}{F} \quad (23)$$

So far, all the coefficients in equations (10-12, 21) are known, and the corresponding jerk, acceleration, velocity and displacement of any time $t$ can be calculated easily.

To demonstrate the advantages of our improved S-shape velocity scheduling function, our velocity scheduling function and velocity scheduling function in [32] are applied to a sub-curve with a start velocity of $0\text{mm/s}$, an end velocity of $20\text{mm/s}$ and an arc length of $50\text{mm}$. The interpolation conditions proposed in this paper are listed in Table 1.

**Table 1** Interpolation conditions proposed in this paper

| **Parameters** | **Symbols** | **Values** |
|---|---|---|
| Command feedrate | $F$ | $200\text{mm/s}$ |
| Maximum tangential acceleration | $A_t$ | $2000\text{mm/s}^2$ |
| Maximum centripetal acceleration | $A_n$ | $2000\text{mm/s}^2$ |
| Maximum tangential jerk | $J_t$ | $60000\text{mm/s}^3$ |
| Maximum centripetal jerk | $J_n$ | $60000\text{mm/s}^3$ |
| Maximum chord error | $\delta$ | $0.001\text{mm}$ |
| Interpolation period | $T_s$ | $0.001\text{s}$ |

As shown in Fig.3, the blue curves are the results of our velocity scheduling function, and the black curves are the results of velocity scheduling function in [32]. It is noted that the feedrate, acceleration and jerk values of both velocity scheduling functions are limited to the maximum values. However, when our velocity scheduling function is applied, the acceleration can maintain at the maximum for a while, and the jerk can reach the maximum. This means that the velocity can be accelerated/decelerated faster from the start/command velocity to the command/end velocity. It can also be proved from the simulation results, When velocity scheduling function in [32] is applied to plan the feed rate, the interpolating time of this curve is $0.43\mathrm{s}$, and the interpolating time is $0.391\mathrm{s}$ if our velocity scheduling function is proposed, which is reduced by $9.07\%$

**a**

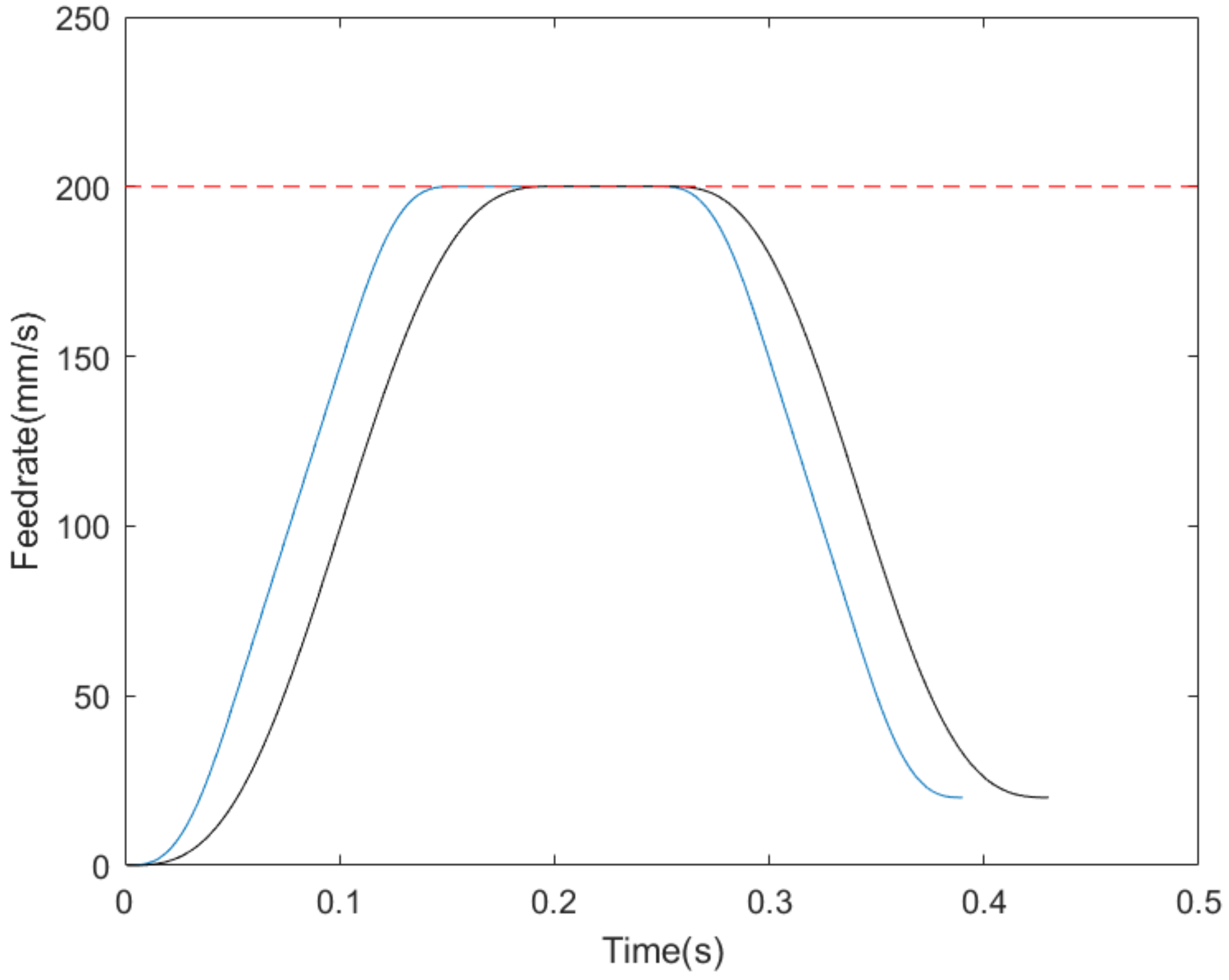

**b**

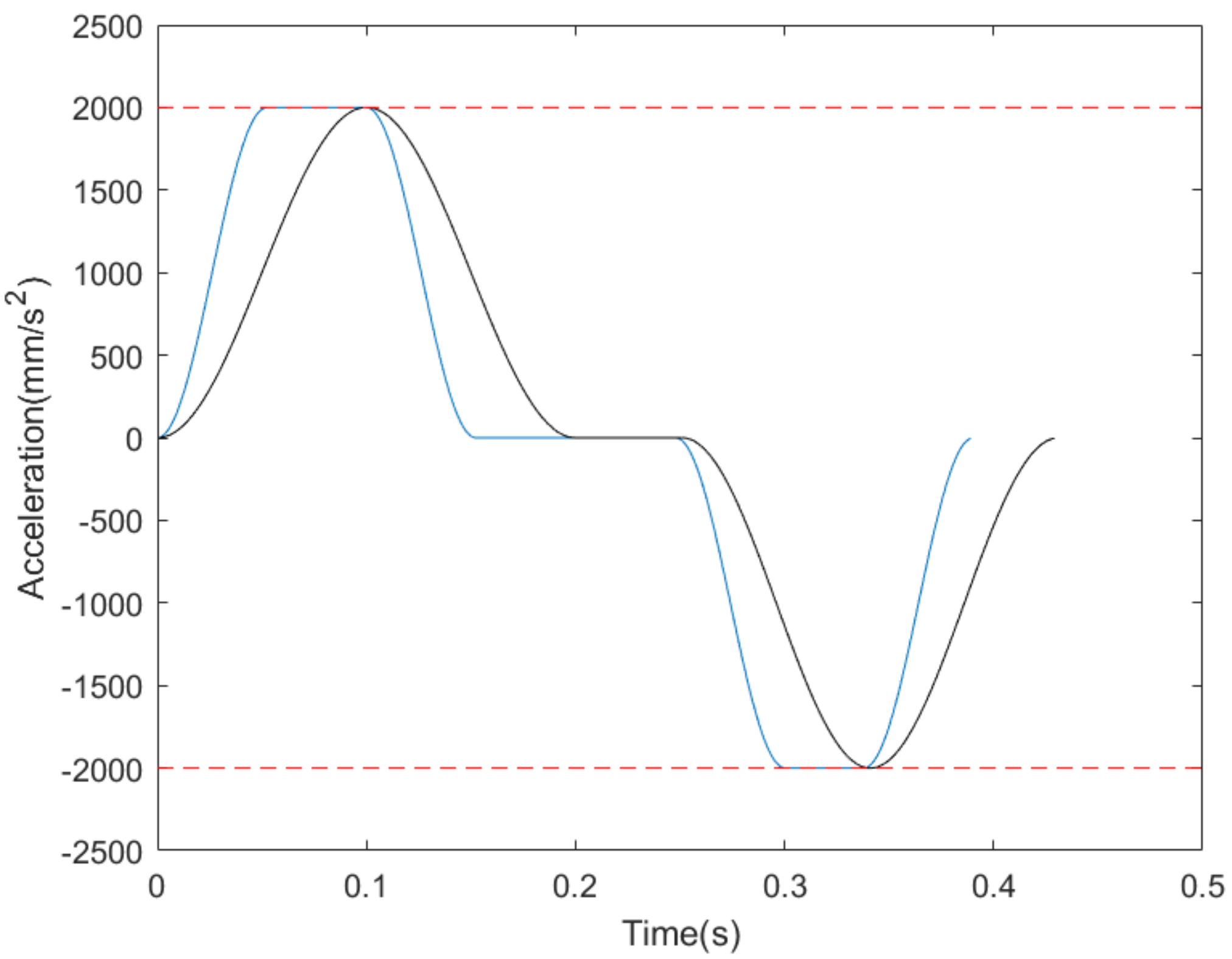

**c**

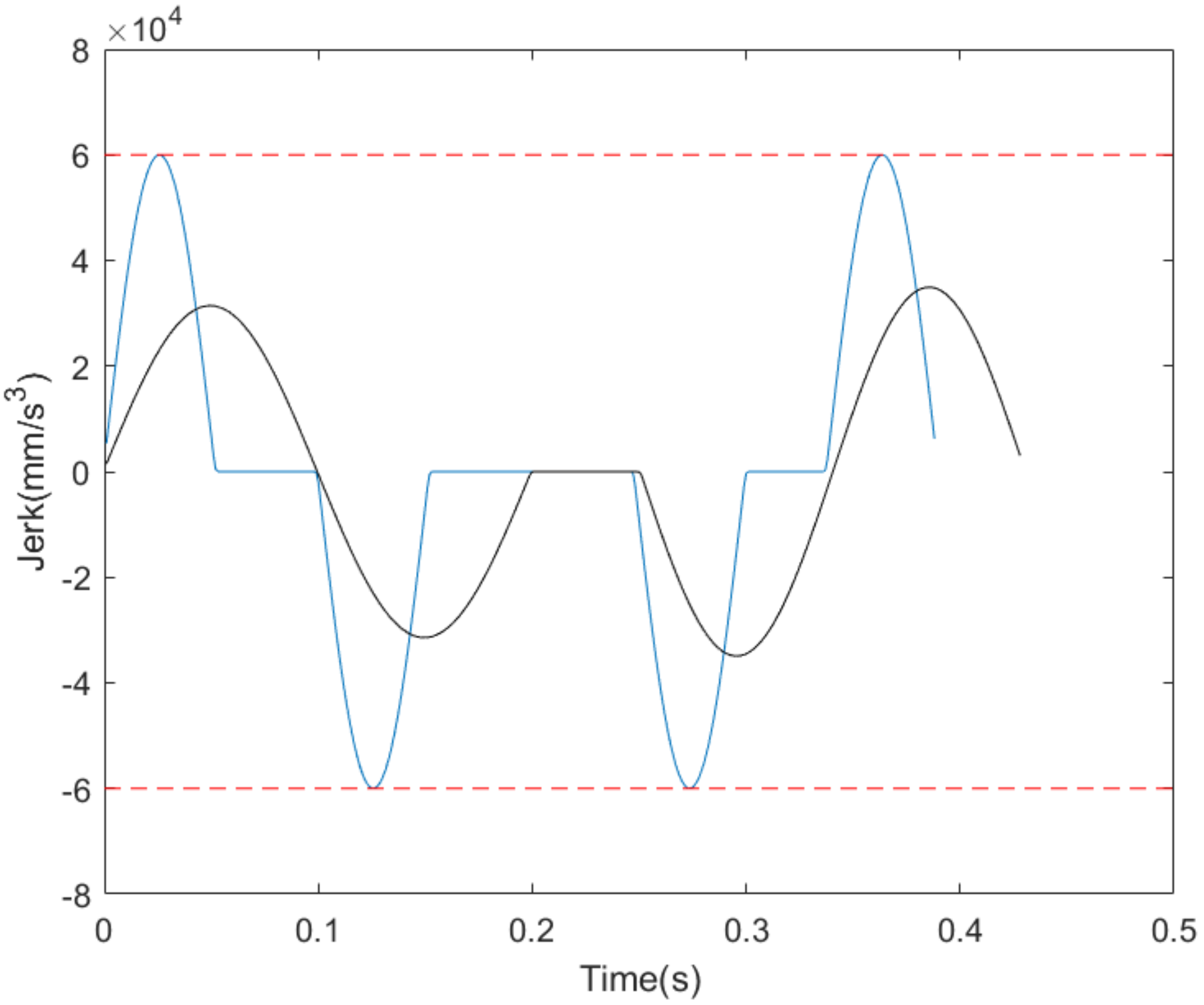

**Fig. 3** Simulation results of a sub-curve by our velocity scheduling function and velocity scheduling function in [32]. **a** Scheduled feedrate. **b** Acceleration. **c** Jerk.

3.2 Bidirectional scanning module

As mentioned above, the start and end velocity of each sub-curve $C_i$ may need to be adjusted. Since the end velocity of sub-curve $C_i$ equals to the start velocity of sub-curve $C_{i+1}$, It is not possible to adjust the endpoint velocity on only one sub-curve, but to adjust the endpoint velocity by treating all sub-curves as a whole.

In Section 2, supposing that there are $n$ breakpoints detected, and the NURBS curve is divided into $n+1$ sub-curves. The maximum allowable feedrate corresponding to each breakpoints is denoted by $v_i(i=1,2,...,n)$, since the federate should be zero at the end points of the NURBS curve, $v_0=0$ and $v_{n+1}=0$ are added into the maximum federate set $\{v_i, i=1,2,...,n\}$. Then the maximum feedrate set $\{v_i, i=0,1,2,...,n,n+1\}$ is adjusted based on velocity scheduling function in Section 3.1, and $v_{i\text{-}1}$ / $v_i$ is assigned to the start/end velocity of sub-curve $C_i(i=1,2,..,n)$.

“Bang-Bang” control is usually adopted in the traditional bidirectional scanning algorithms, the major concern of which is to reach the acceleration or torque limit of one axis at any time[39]. However, the resulting acceleration and jerk profiles are discontinuous. In this paper, the jerk continuity and acceleration continuity are both considered in bidirectional scanning module. Considering the kinematic capabilities, the three-phases S-shape acceleration in Section 3.1 is proposed in both the forward and backward scanning. The proposed bidirectional scanning module consisting of two steps, the flowchart of STEP I is shown as Fig.4, and the details are summarized as follows.

STEP I

In the first step, forward scanning is performed from $v_0=0$ in the maximum federate set $\{v_i, i=0,1,2,...,n,n+1\}$. The starting velocity is set to be zero, so the tool is guaranteed to start moving from rest at the beginning of the NURBS curve. Since the chord error, centripetal acceleration, centripetal jerk and feed rate limitations are considered in Section 2, only tangential acceleration and tangential jerk need to be considered in this step. Intersecting these constraints, if $v_{i+1}>v_i$, the acceleration is performed to obtain a kinematic maximum allowable feedrate $v$, then $v_{i+1}$ is updated to $v$ if $v_{i+1}>v$. This process is performed recursively until the end of the NURBS curve is reached. The detailed algorithm is described as follows.

(1) Let $i=0$.

(2) If $v_{i+1}>v_i$, go to (3); otherwise, go to (6).

(3) If $v_{i+1}-v_i \le \frac{\pi A_t^2}{2J_t}$, go to (4); otherwise, go to (5).

(4) Let $T_1=T_3=\sqrt{\frac{\pi(v_{i+1}-v_i)}{2J_t}}$ and $T_2=0$, calculate

$$s=(v_i+v_{i+1})(2T_1+T_2)/2=(v_i+v_{i+1})\sqrt{\frac{\pi(v_{i+1}-v_i)}{2J_t}} \tag{24}$$

If $s<l_i$, $v_{i+1}$ does not need to be adjusted; otherwise, the actually maximum velocity $v$ has the following relationship with $v_i$ and $l_i$:

$$(v_i + v)\sqrt{\frac{v - v_i}{2J_t}} = l_i \tag{25}$$

Equation (25) can be transformed to a univariate cubic equation:

$$v^3 + v_i v^2 - v_i^2 v - v_i^3 - \frac{2l_i^2 J_t}{\pi} = 0 \tag{26}$$

It can be proved that equation (26) has one and only root $v_{update}$ satisfying: $v_i < v_{update} < v_{i+1}$, then $v_{i+1}$ is updated to $v_{update}$, go to (6).

(5) Let $T_1 = T_3 = \frac{\pi A_t}{2J_t}$ and $T_2 = \frac{v_{i+1} - v_i}{A_t} - T_1$, calculate

$$s = (v_i + v_{i+1})(2T_1 + T_2)/2 = (v_i + v_{i+1})(\frac{v_{i+1} - v_i}{A_t} + \frac{\pi A_t}{2J_t}) \tag{27}$$

If $s < l_i$, $v_{i+1}$ does not need to be adjusted; otherwise, the relationship between actually maximum velocity $v$, $v_i$ and $l_i$ has two condition.

(5a) The actual maximum velocity $v$ satisfying: $v - v_i > \frac{\pi A_t^2}{2J_t}$. Then

$$(v + v_i)(\frac{v - v_i}{A_t} + \frac{\pi A_t}{2J_t}) = 2l_i \tag{28}$$

Equation (28) can be transformed to a univariate quadratic equation:

$$\frac{v^2}{2A_t} - \frac{\pi A_t v}{4J_t} + \frac{\pi A_t v_i}{4J_t} - \frac{v_i^2}{2A_t} - l_i = 0 \tag{29}$$

(5b) The actual maximum velocity $v$ satisfying: $v - v_i \le \frac{\pi A_t^2}{2J_t}$. The relationship is the same as equation (25) and (26).

It can be proved that there is one and only root $v_{update}$ satisfying condition (5a) or (5b). then $v_{i+1}$ is updated to $v_{update}$, go to (6).

(6) Let $i = i + 1$, if $i \leq n$, go to (2); otherwise, stop.

STEP II

In the second step, backward scanning is performed from $v_{n+1}=0$ in the maximum federate set $\{v_i, i = 0,1,2,...,n,n+1\}$. The end velocity is set to be zero, and it will not be updated in STEP I, so the tool is guaranteed to stop at the end of the NURBS curve. The detailed algorithm proposed is similar to STEP I, only minor changes are required as follows: (1) $i=0$ is replaced by $i=n+1$. (2) $v_i$ and $v_{i+1}$ interchange. (3) $i=i+1$ is replaced by $i=i-1$. (4) $i \leq n$ is replaced by $i \geq 1$.

After these two steps, it can be guaranteed that on each sub-curve $C_i$, the feedrate can be changed from the start velocity $v_{start} = v_{i-1}$ to the end velocity $v_{end} = v_i$ when moving to the end point.

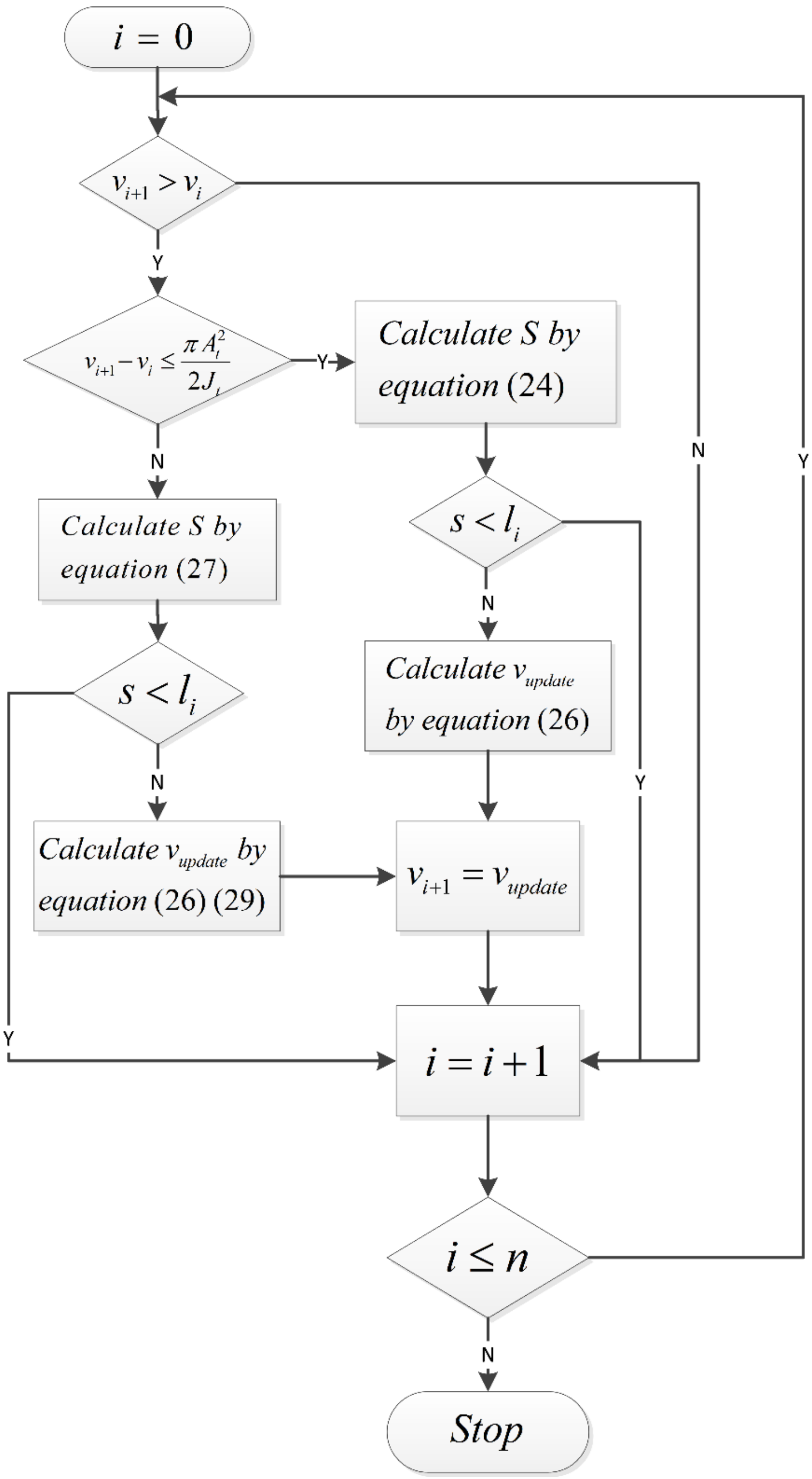

**Fig.4** The flowchart of STEP I

3.3 Velocity scheduling module

After the process of bidirectional scanning module, the sub-curve may be two situations of the path: the case of acceleration/deceleration (ACC-or-DEC case) and the case of acceleration–uniform–deceleration or acceleration–deceleration (ACC-and-DEC case). For ACC-or-DEC case, the feed rate of at least one endpoint of the sub-curve has been adjusted by bidirectional scanning module. The actual maximum velocity must be the start velocity or the end velocity, and the feed rate keeps accelerating or decelerating along the sub-curve. For ACC-and-DEC case, the feed rate at any endpoint of the sub-curve is not adjusted. If the sub-curve is long enough, the feed rate can reach the command velocity $F$. The feed rate accelerates from the start velocity $v_{start}$ to the command velocity $F$, maintains the velocity $F$ for a period of time and finally decelerates to the end velocity $v_{end}$. If the sub-curve is not long enough, the feed rate can not reach the command velocity $F$. The feed rate accelerates from the start velocity $v_{start}$ to the actual maximum velocity $v_m$ $(v_m < F)$, then decelerates to the end velocity $v_{end}$ directly. For different cases, the processing details of the velocity scheduling module are as follows.

3.3.1 ACC-or-DEC case

Supposing that the start velocity $v_{start}$ is lower than the end velocity $v_{end}$, the feed rate keeps accelerating along the sub-curve. Since there is no uniform and decelerating period, it is obviously that $T_4 = T_5 = T_6 = T_7 = 0$.

According to the basic model in Section 3.1, if $v_{end} - v_{start} \leq \frac{\pi A_t^2}{2J_t}$, the acceleration of the feed rate first increases from zero to the actual maximum acceleration $A_1$ and then decreases to zero, which implies that $T_2 = 0$. According to equation (17) and (21), we can obtain that:

$$(v_{end} - v_{start}) / T_1 = A_1 \tag{30}$$

$$(v_{start} + v_{end}) T_1 = l_i \tag{31}$$

Solve the above equation (30-31), the actual maximum acceleration $A_1$ and time interval $T_1$ and $T_3$ can be calculated as follows:

$$A_1 = \frac{v_{end}^2 - v_{start}^2}{l_i} \tag{32}$$

$$T_1 = T_3 = \frac{l_i}{v_{end} + v_{start}} \tag{33}$$

If $v_{start} - v_{end} > \frac{\pi A_t^2}{2J_t}$, the acceleration of the feed rate first increases from zero to the tangential acceleration limitation $A_t$, then keeps the acceleration constant for a period of time and decrease to zero at last. From equation (20) it is obvisouly that $A_1 = A_t$, $T_1 = T_3 = \frac{\pi A_t}{2J_t}$ and $T_2 = \frac{v_{end} - v_{start}}{A_t} - T_1$.

For the situation that the start velocity $v_{start}$ is higher than the end velocity $v_{end}$, the feed rate keeps decelerating along the sub-curve. It is obviously that $T_1 = T_2 = T_3 = T_4 = 0$, and the calculation process of the maximum acceleration during the decelerating period and other time intervals $T_5$, $T_6$ and $T_7$ is similar to the previous paragraph.

3.3.2 ACC-and-DEC case

For this case, the key question is whether the actual maximum velocity $v_m$ can reach the command velocity $F$. This can be judged by solving equation (23) in the basic model in Section 3.1. If $T_4 \geq 0$, the actual maximum velocity $v_m$ can reach the command velocity $F$, and all time intervals and actual maximum in acceleration and deceleration can be obtained easily. If $T_4 < 0$, the actual maximum velocity $v_m$ can't reach the command velocity $F$, the feed rate accelerates from the start velocity $v_{start}$ to the actual maximum velocity $v_m$ , then decelerates to the end velocity $v_{end}$, where $v_m$ satisfying $v_m > \max(v_{start}, v_{end})$ and $v_m < F$. Assuming $v_m$ is known, then the profile of feed rate is consisting of an acceleration and a deceleration phases, and the time intervals and the actual maximum acceleration and deceleration can be obtained by similar steps as in Section 3.3.1.

The only variable that needs to be solved is the actual maximum velocity $v_m$ now. For simplicity, assuming that $v_{start} > v_{end}$, and the process is similar when $v_{start} \leq v_{end}$. The detailed process of solving the actual maximum velocity is as follows.

(1) Let $T_1 = T_3 = \frac{\pi A_t}{2J_t} = T_5 = T_7$ , $T_2 = \frac{v_m - v_{start}}{A_t} - \frac{\pi A_t}{2J_t}$ and $T_6 = \frac{v_m - v_{end}}{A_t} - \frac{\pi A_t}{2J_t}$ .

According to equation (20), the displacement can be expressed by

$$s = \frac{(v_m + v_{start})(2T_1 + T_2) + (v_m + v_{end})(2T_5 + T_6)}{2} \tag{34}$$

To guarantee the displacement equals to the arc length $l_i$ , according to equation (34), an univariate quadratic equation can be obtained as follows:

$$\frac{2}{A_t} v_m^2 + \frac{\pi A_t}{J_t} v_m + \frac{\pi A_t}{2J_t}(v_{start} + v_{end}) - \frac{v_{start}^2 + v_{end}^2}{A_t} - 2l_i = 0 \tag{35}$$

If there exits one root $v_{m,j}$ satisfying

$$v_{start}+\frac{\pi A_t^2}{2J_t}<v_{m,j}<F \tag{36}$$

then $v_m = v_{m,j}$ , go to (4); otherwise, go to (2).

(2) Let $T_1$=$T_3$=$\sqrt{\frac{\pi(v_m - v_{start})}{2J_t}}$ and $T_2$=0 . To guarantee the displacement equals to the arc length $l_i$ , according to equation (34), an univariate quartic equation can be obtained as follows:

$$\begin{aligned}&\frac{1}{4A_t^2}v_m^4-\frac{\pi}{4J_t}v_m^3+[\frac{\pi^2A_t^2}{16J_t^2}-\frac{\pi v_s}{2J_t}+\frac{1}{A_t}(\frac{\pi A_t v_e}{4J_t}-\frac{v_e^2}{2A_t}-l_i)]v_m^2\\&+[\frac{\pi v_s^2}{2J_t}+\frac{\pi A_t}{2J_t}(\frac{\pi A_t v_e}{4J_t}-\frac{v_e^2}{2A_t}-l_i)]v_m+\frac{\pi v_s^3}{2J_t}+(\frac{\pi A_t v_e}{4J_t}-\frac{v_e^2}{2A_t}-l_i)^2=0\end{aligned} \tag{37}$$

If there exits one root $v_{m,j}$ satisfying:

$$\max(v_{start},v_{end}+\frac{\pi A_t^2}{2J_t})<v_m<\min(v_{start}+\frac{\pi A_t^2}{2J_t},F) \tag{38}$$

then $v_m = v_{m,j}$ , go to (4); otherwise, go to (3).

(3) Let $T_5$=$T_7$=$\sqrt{\frac{\pi(v_m - v_{end})}{2J_t}}$ and $T_6$=0 , To guarantee the displacement equals to the arc length $l_i$ , according to equation (34), an univariate quartic equation can be obtained as follows:

$$\begin{aligned}&(v_e-v_s)^2v_m^4+[2(v_e-v_s)(v_s^2-v_e^2)-\frac{8J_tl_i^2}{\pi}]v_m^3\\&+[(v_s^2-v_e^2)^2+2(v_e-v_s)(v_s^3-v_e^3+\frac{2J_tl_i^2}{\pi})-\frac{8J_tl_i^2}{\pi}v_e]v_m^2\\&+[2(v_s^2-v_e^2)(v_s^3-v_e^3+\frac{2J_tl_i^2}{\pi})+\frac{8J_tl_i^2}{\pi}v_e^2]v_m\\&+(v_s^3-v_e^3+\frac{2J_tl_i^2}{\pi})^2+\frac{8J_tl_i^2}{\pi}v_e^3=0\end{aligned} \tag{39}$$

It can be proven that there exits one and only root $v_{m,j}$ satisfying:

$$\mathrm{v}_{start} < v_m < \min(v_{end} + \frac{\pi A_t^2}{2J_t}, F) \quad (40)$$

then $v_m = v_{m,j}$ , go to (4).

(4) Stop.

3.4 Round-off error elimination module

In most cases, the total interpolating time of the NURBS curve is not an integer multiple of the interpolation period, which leads to round-off error. To eliminate the round-off error, the common solution is to allocate the displacement corresponding to the time of less than one interpolation period to other interpolation periods according to a certain function. However, these solutions will lead to the excess of chord error, feed rate, acceleration or jerk in some points since the displacement of all interpolation periods are increased.

In this paper, the total interpolating time $T_{total} = \sum_{i=1}^{n+1} \sum_{j=1}^{7} T_{i,j}$ is increased to $T_{total} + \Delta t$ , where the extended time $\Delta t$ is calculated as follows:

$$\Delta t = \lceil T_{total} / T_s \rceil * T_s - T_{total} \quad (41)$$

It is noted that $\Delta t$ is less than an interpolation period $T_s$ and $T_{total} + \Delta t$ is an integer multiple of the interpolation period. In order to reduce the amount of calculation, only the feed rate profile of one sub-curve changes, which requires that the start velocity and the end velocity of this sub-curve do not change. For sub-curve of ACC-or-DEC case, as $(v_{start} + v_{end})T = l_i$ , the interpolating time of this sub-curve can't change without changing the start velocity, the end velocity or the arc-length. For sub-curve of ACC-and-DEC case, if the actual maximum velocity $v_m$ is reduced appropriately, the interpolating time of this curve can be increased by $\Delta t$ without changing the start velocity or end velocity.

For simplicity, the first sub-curve of ACC-and-DEC case is chosen to execute round-off error elimination module, and assuming that $v_{start} > v_{end}$ .(the process is similar when $v_{start} \le v_{end}$ ) Denoting the interpolating time of this sub-curve as $T = \sum_{i=1}^{n} T_i$ , where the time intervals $T_i \ (i = 1, 2, ..., 7)$ are obtained by velocity scheduling module. To increase the interpolating time from $T$ to $T + \Delta t$, the actual maximum velocity is reduced from $v_m$ to $v_m^{'}$. The detailed process of calculating the actual maximum velocity is as follows.

(1) Let $T_1 = T_3 = \frac{\pi A_t}{2J_t} = T_5 = T_7$, $T_2 = \frac{v_m^{'} - v_{start}}{A_t} - \frac{\pi A_t}{2J_t}$ and $T_6 = \frac{v_m^{'} - v_{end}}{A_t} - \frac{\pi A_t}{2J_t}$. According to equation (20), the displacement can be expressed by

$$s = \frac{(v_m^{'} + v_{start})(2T_1 + T_2) + (v_m^{'} + v_{end})(2T_5 + T_6)}{2} + v_m^{'}(T - 2T_1 - T_2 - 2T_5 - T_7) \quad (42)$$

To guarantee the displacement equals to the arc length $l_i$ , according to equation (42), an univariate quadratic equation can be obtained as follows:

$$-\frac{1}{A_t} v_m^{'2} + (T + \frac{v_{start}}{A_t} + \frac{v_{end}}{A_t} - \frac{\pi A_t}{2J_t}) v_m^{'} + \frac{\pi A_t}{4J_t}(v_{start} + v_{end}) - \frac{v_{start}^2 + v_{end}^2}{2A_t} - l_i = 0 \quad (43)$$

If there exits one root $v'_{m,j}$ satisfying

$$v_{start} + \frac{\pi A_t^2}{2J_t} < v'_{m,j} < v_m \tag{44}$$

then $v'_m = v'_{m,j}$ , go to (4); otherwise, go to (2).

(2) Let $T_1 = T_3 = \sqrt{\frac{\pi(v'_m - v_{start})}{2J_t}}$ and $T_2 = 0$, To guarantee the displacement equals to the arc length $l_i$ , according to equation (42), an univariate quartic equation can be obtained as follows:

$$\begin{aligned} &\frac{1}{4A_t^2} v'^4_m + \left(\frac{a}{A_t} - \frac{\pi}{2J_t}\right) v'^3_m + \left(a^2 + \frac{b}{A_t} + \frac{3v_s \pi}{2J_t}\right) v'^2_m \\ &+ \left(2ab - \frac{3v_s^2 \pi}{2J_t}\right) v'_m + b^2 + \frac{3v_s^3 \pi}{2J_t} = 0 \end{aligned} \tag{45}$$

Where $a = \frac{\pi A_t}{4J_t} - T_s - \frac{v_e}{A_t}$ and $b = l_i + \frac{v_e^2}{2A_t} - \frac{\pi A_t v_e}{4J_t}$ . If there exits one root $v'_{m,j}$ satisfying:

$$\max\left(v_{start}, v_{end} + \frac{\pi A_t^2}{2J_t}\right) < v'_m < \min\left(v_{start} + \frac{\pi A_t^2}{2J_t}, v_m\right) \tag{46}$$

then $v'_m = v'_{m,j}$ , go to (4); otherwise, go to (3).

(3) Let $T_5 = T_7 = \sqrt{\frac{\pi(v'_m - v_{end})}{2J_t}}$ and $T_6 = 0$, To guarantee the displacement equals to the arc length $l_i$ , according to equation (42), an univariate quartic equation can be obtained as follows:

$$
\begin{aligned}
&T_s^2 v_m^{'5} - (2l_i T_s + 3v_e T_s^2 + \frac{a^2 J_t}{2\pi}) v_m^{'4} + (l_i^2 + 6l_i T_s v_e + 3v_e^2 T_s^2 - \frac{abJ_t}{\pi}) v_m^{'3} \\
&-(3v_e l_i^2 + 6l_i T_s v_e^2 + v_e^3 T_s^2 + \frac{b^2 J_t}{2\pi} + \frac{acJ_t}{\pi}) v_m^{'2} \\
&+(3v_e^2 l_i^2 + 2l_i T_s v_e^3 - \frac{bcJ_t}{\pi}) v_m^{'} - (l_i^2 v_e^3 + \frac{c^2 J_t}{2\pi}) = 0
\end{aligned}
\tag{47}
$$

Where $a = \frac{3\pi(v_e - v_s)}{2J_t} - T_s^2$ , $b = \frac{3\pi(v_s^2 - v_e^2)}{2J_t} + 2l_i T_s$ and $c = \frac{\pi(v_e^3 - v_s^3)}{2J_t} - l_i^2$ . It can be proven that there exits one and only root $v_{m,j}^{'}$ satisfying:

$$
\mathrm{v}_{start} < v_m^{'} < \min(v_{end} + \frac{\pi A_t^2}{2J_t}, v_m) \tag{48}
$$

then $v_m^{'} = v_{m,j}^{'}$ , go to (4).

(4) Stop.

To demonstrate the effect of our round-off error elimination module, our round-off error elimination module and tradition round-off error compensation module are applied to the same sub-curve in section 3.1. As shown in Fig.5.a, the blue curve is the feedrate with tradition round-off error compensation module, and the black curve is the feedrate with our round-off error elimination module. From the enlarged region 1 in Fig.5.b, It is noted that the blue curve is a little above the red dotted line, which means the actual velocity exceeds the given maximum value, while the black curve is below the red dotted line strictly. From the enlarged region 2 in Fig.5.c, We know that the interpolating of this sub-curve with tradition round-off error compensation module is $0.39s$ , and the interpolating time with our round-off error elimination module is $0.391s$ . Although the interpolating time is increased by an interpolation cycle, the velocity is strictly limited to the given maximum value, and this time loss is worthwhile.

**a**

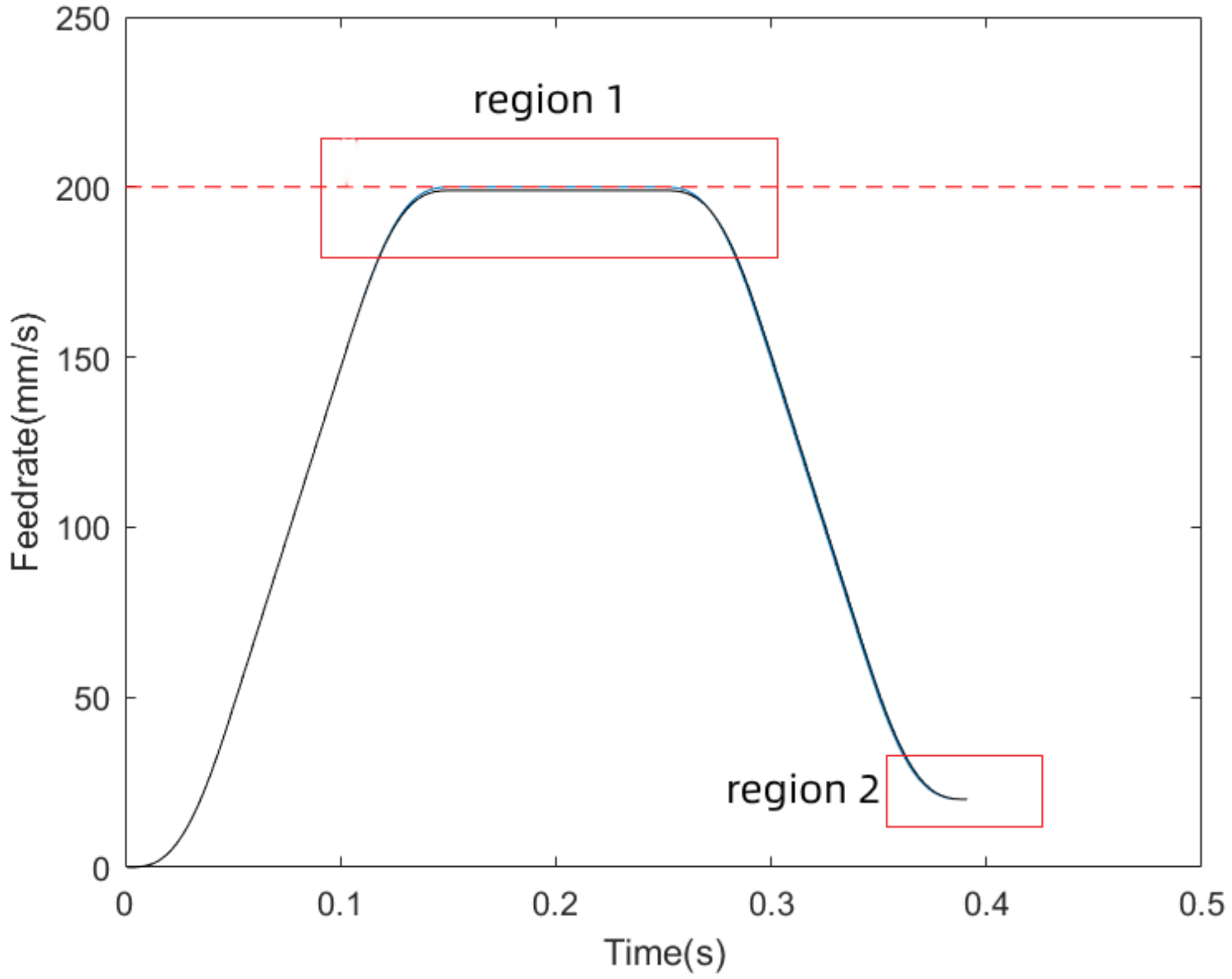

**b**

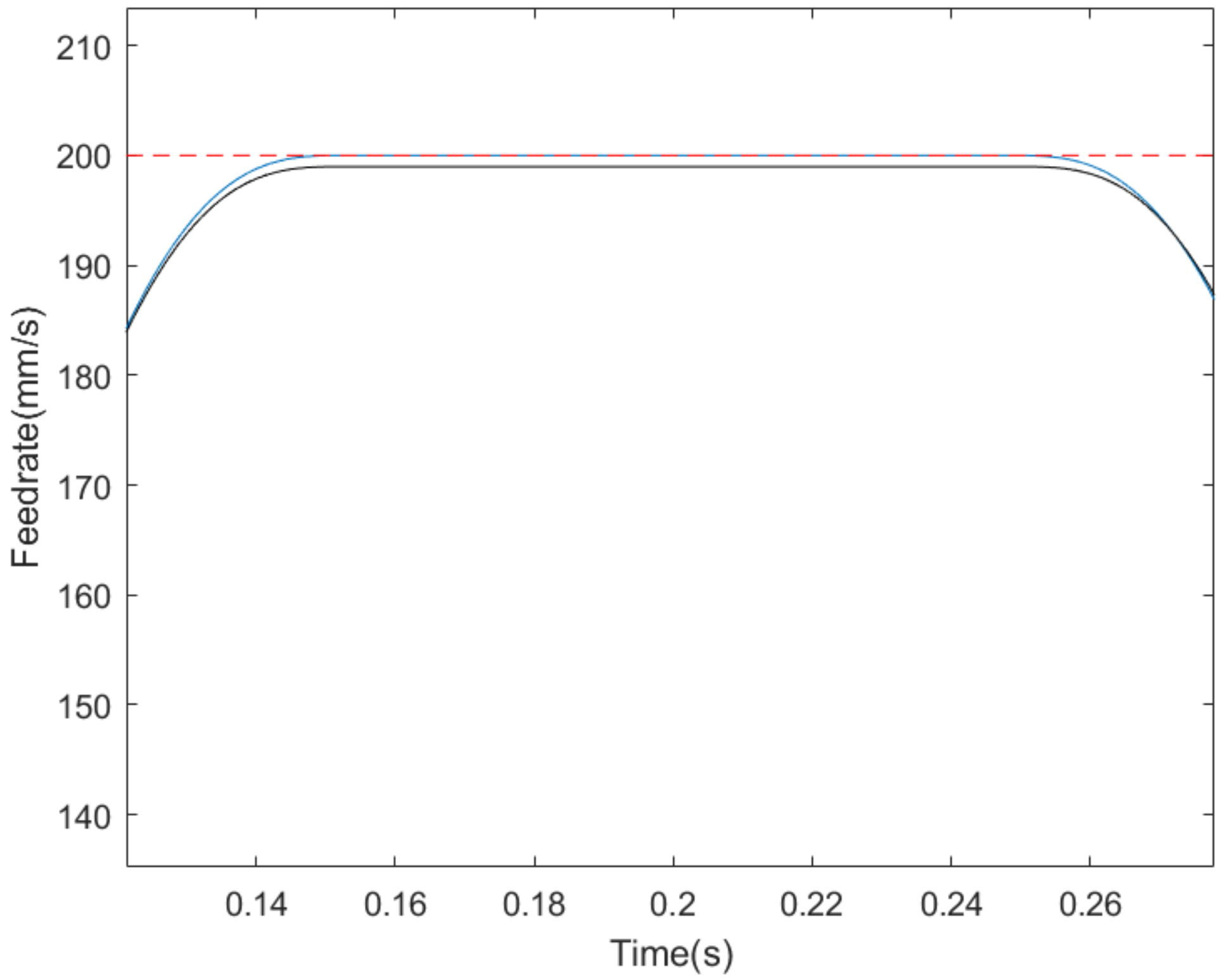

**c**

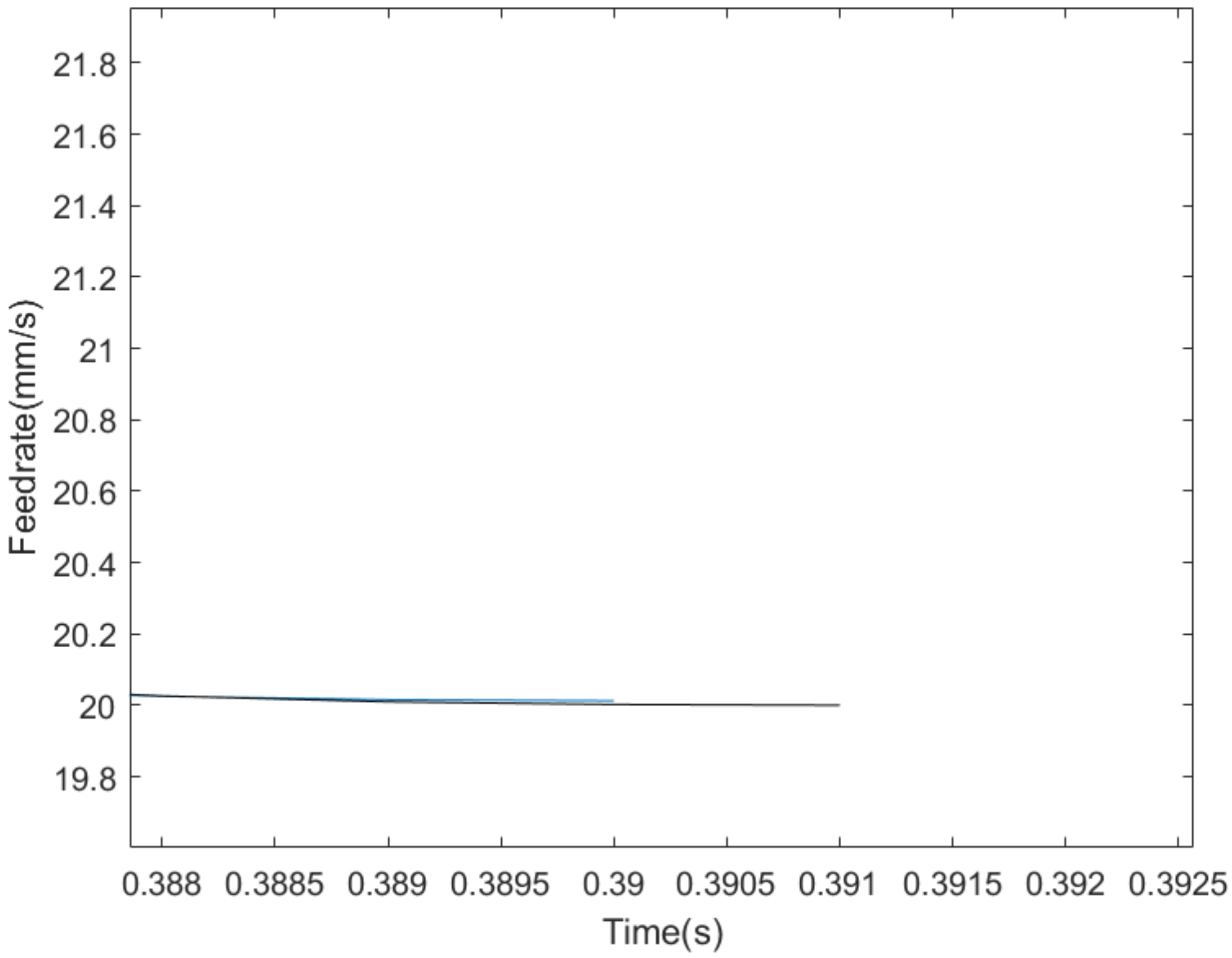

**Fig. 5** Simulation results of a sub-curve with our round-off error elimination module

and tradition round-off error compensation module. **a** Scheduled feedrate. **b** the enlarged region 1. **c** the enlarged region 2.

3.5 summary of the proposed method

To give a clear illustration of the proposed feedrate scheduling method, the flowchart of the proposed method is summarized as follows. First, the parameters of the NURBS and the initial constraints including chord error and dynamic constraints of feedrate scheduling are given. Then, the breakpoints of the NURBS are detected by equations (6-9), and it is divided to several sub-curves according to the breakpoints. And the maximum allowable feedrate corresponding to each breakpoints are calculated by STEP I and STEP II in section 3.2, which guarantees that the continuity of the feedrate at the junctions between successive NURBS sub-curves. Later, for each sub-curve $C_i$, the time intervals $T_{i,j}(j=1,2,...,7)$, the actual maximum acceleration $A_{i,1}$ and the actual maximum deceleration $A_{i,2}$ are calculated following the process in section 3.3. After the extended time $\Delta t$ is calculated by equation (41), the time intervals $T_{k,j}(j=1,2,...,7)$, the actual maximum acceleration $A_{k,1}$ and the actual maximum deceleration $A_{k,2}$ of the first sub-curve with ACC-and-DEC case (whose index equals to $k$) are recalculated according to the process in section 3.4. Finally, for each time $t$, the displacement along the NURBS can be calculated according to equation (20).

## 4. Simulation results and analysis

In this section, two typical NURBS curves named butterfly-shaped curve and trident curve, shown in Fig. 6, are conducted as examples to evaluate the performance of the proposed feedrate scheduling method. The degrees, control points, knot vectors, and weight vectors of the test curves are given in Appendix.

The interpolation conditions proposed in this paper are listed in Table 1. Based on the interpolation conditions, the butterfly-shaped curve, the trident curve and their break points are shown in Fig. 6.

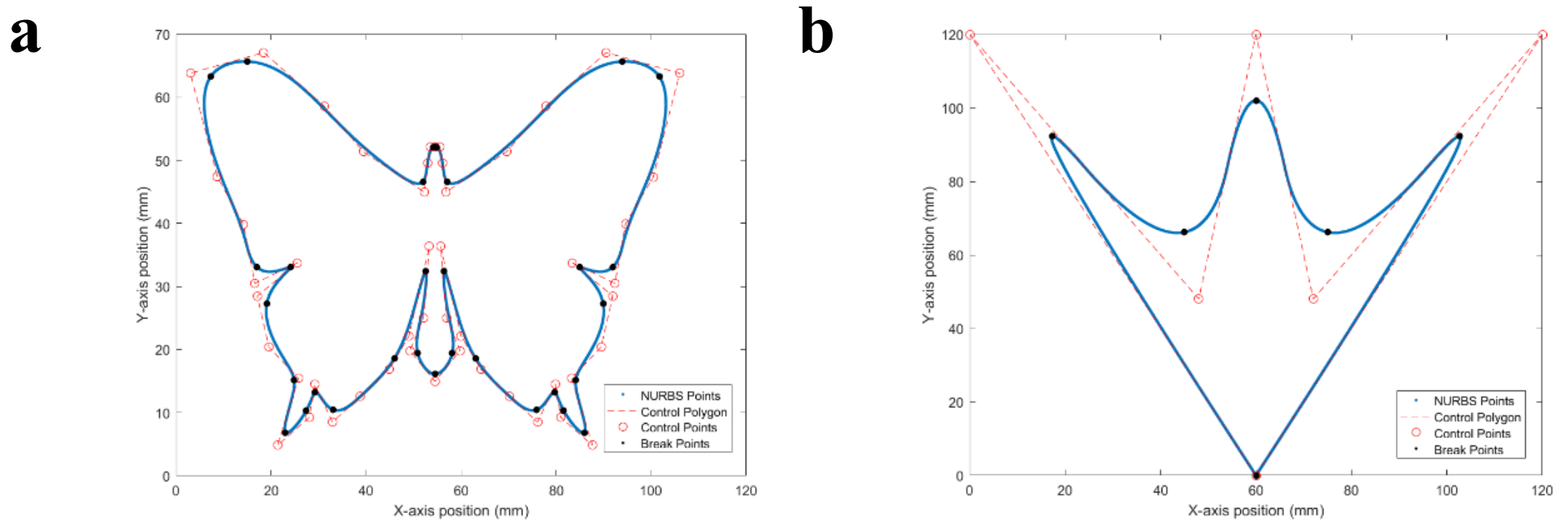

**Fig. 6** Test curves. **a** Butterfly-shaped curve. **b** Trident curve

4.1 Analysis and comparisons of butterfly-shaped curve

As shown in Fig.6 a, there are 32 break points on the NURBS curve, since it is a closed curve, the butterfly-shaped curve is divided into 32 segments. The simulation results obtained through the proposed feedrate scheduling method are shown in Fig.7. From Fig.7 a-c, we know that the feedrate, acceleration and jerk values of proposed interpolation algorithm are limited to the maximum values. Meanwhile, the acceleration and jerk profile is continuous, which means the feedrate profile is smooth enough. In Fig.7 d, the red curve represents the maximum allowable feedrate obtained by equation (9), and the blue curve represents the actual feedrate obtained by the proposed method. We can notice that the blue curve is always below the red curve, which means that the actual feedrate always does not excess the maximum allowable feedrate. As shown in Fig.7 e, the chord errors of all interpolation points are less than $2.5\times10^{-4}\text{mm}$, which is far less than the given threshold $\delta=1\times10^{-3}mm$. This demonstrates that the proposed method has high accuracy.

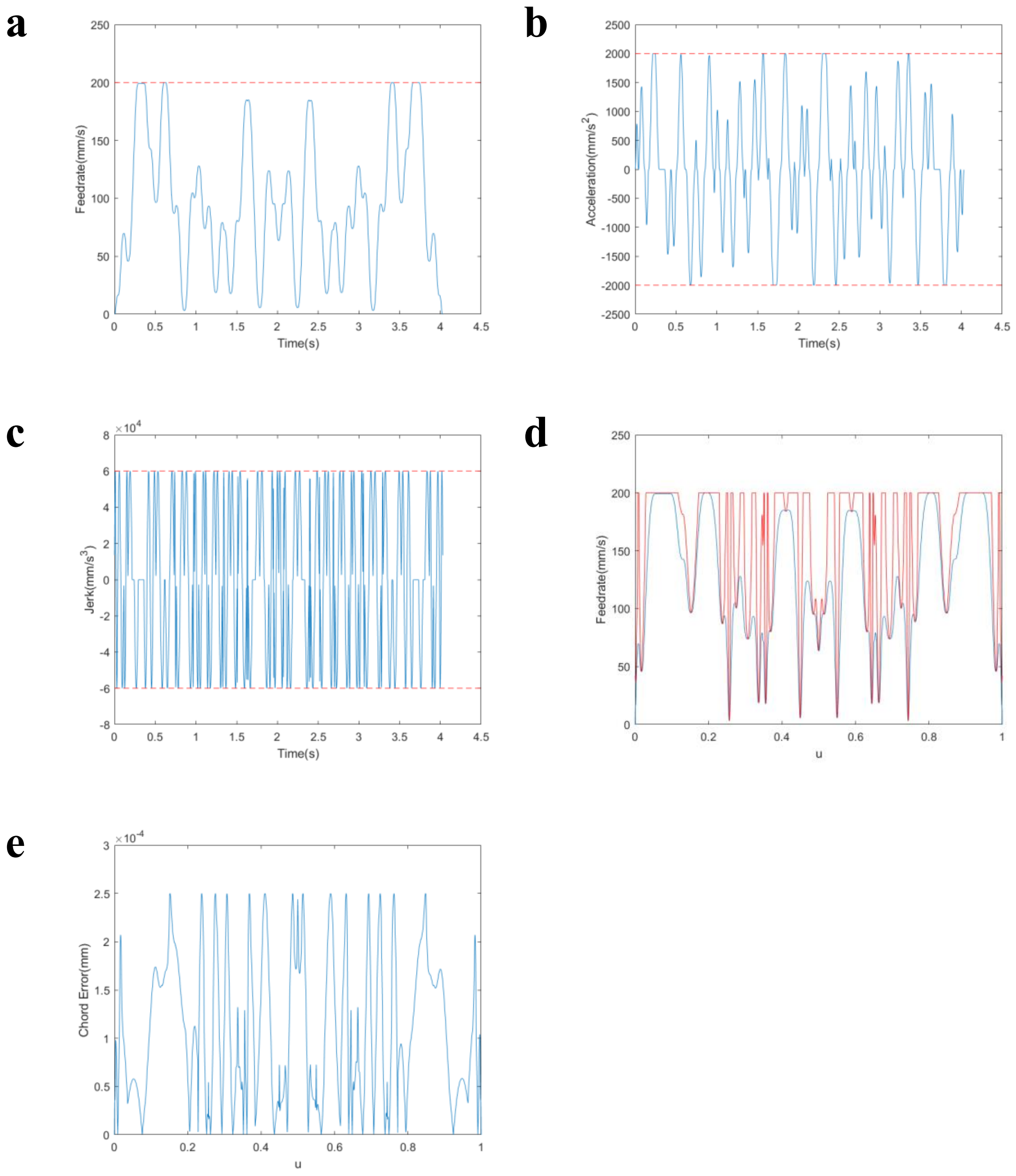

**Fig. 7** Simulation results of butterfly-shaped curve by proposed method. **a** Scheduled feedrate. **b** Acceleration. **c** Jerk. **d** Comparison of actual feedrate and maximum

allowable feedrate. **e** Chord error.

Since some segments of the butterfly-shaped curve are not long enough to have a constant feedrate section, the methods chosen for comparison should be able to deal with short segments. The methods in [25, 36] are chosen, and the simulation results are shown in Fig.8 and Fig.9, respectively. As can be seen, the federate and acceleration profile in Fig.9 a-b slightly excess the confined range in some interpolating points, this is because the round-off error compensation allocates a short displacement to other interpolation periods according to a trapezoidal ACC/DEC algorithm [36]. As for the jerk profile, although both the jerk profile in Fig.8 c and Fig. 9 c excess the confined range in some interpolating points, the jerk profile in Fig.8 c still keeps in a certain extent, while the jerk profile in Fig.9 c exceed a lot. Meanwhile, both the jerk profile in Fig.8 c and Fig. 9 c are discontinuous. Therefore, it will lead to the excessive vibration of the machine tool, and the interpolating quality and efficiency will be affected. In Fig. 9 d, part of the blue curve is above the red curve, which means that the actual feedrate excess the maximum allowable feedrate in some interpolating points, then the motion smoothness will be affected. From Table 2, compared with methods in [25] and [36], the proposed method reduces the average chord error by 13.13% and 13.53%, respectively. And the proposed method has the smallest maximum chord error as well as the actual maximum feedrate, acceleration and jerk. However, The total interpolation time of the proposed method is longer than two other methods by 8.25% and 8.57%, respectively. The main reason is that in order to achieve continuity of the jerk profile, the ‘Bang-Bang Control’ is not satisfied, so the increase of interpolation time is inevitable.

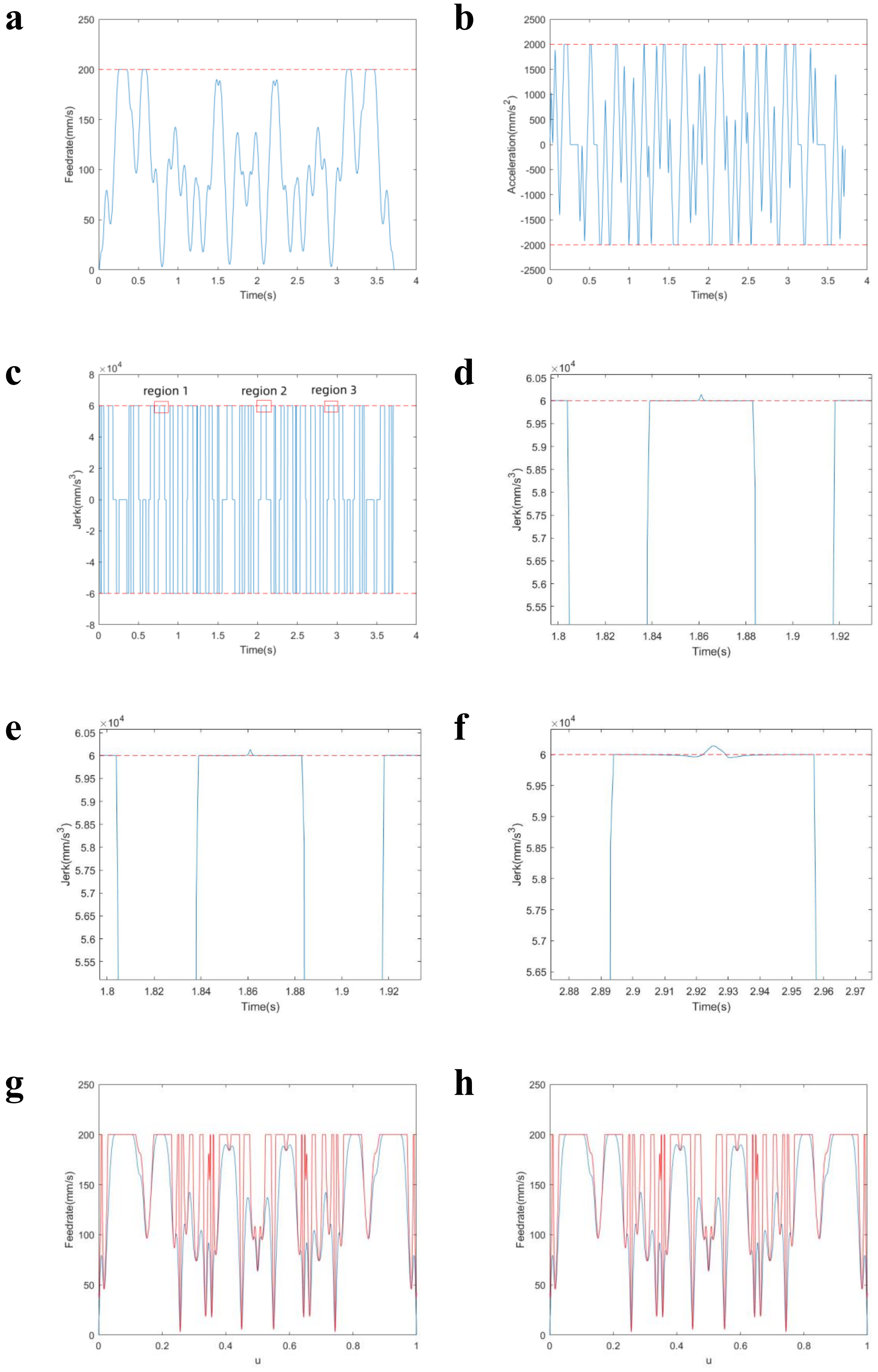

**Fig. 8** Simulation results of butterfly-shaped curve by method in [25]. **a** Scheduled feedrate. **b** Acceleration. **c** The enlarged region 1. **d** The enlarged region 2. **e** The enlarged region 3. **f** Jerk. **g** Comparison of actual feedrate and maximum allowable feedrate. **h** Chord error.

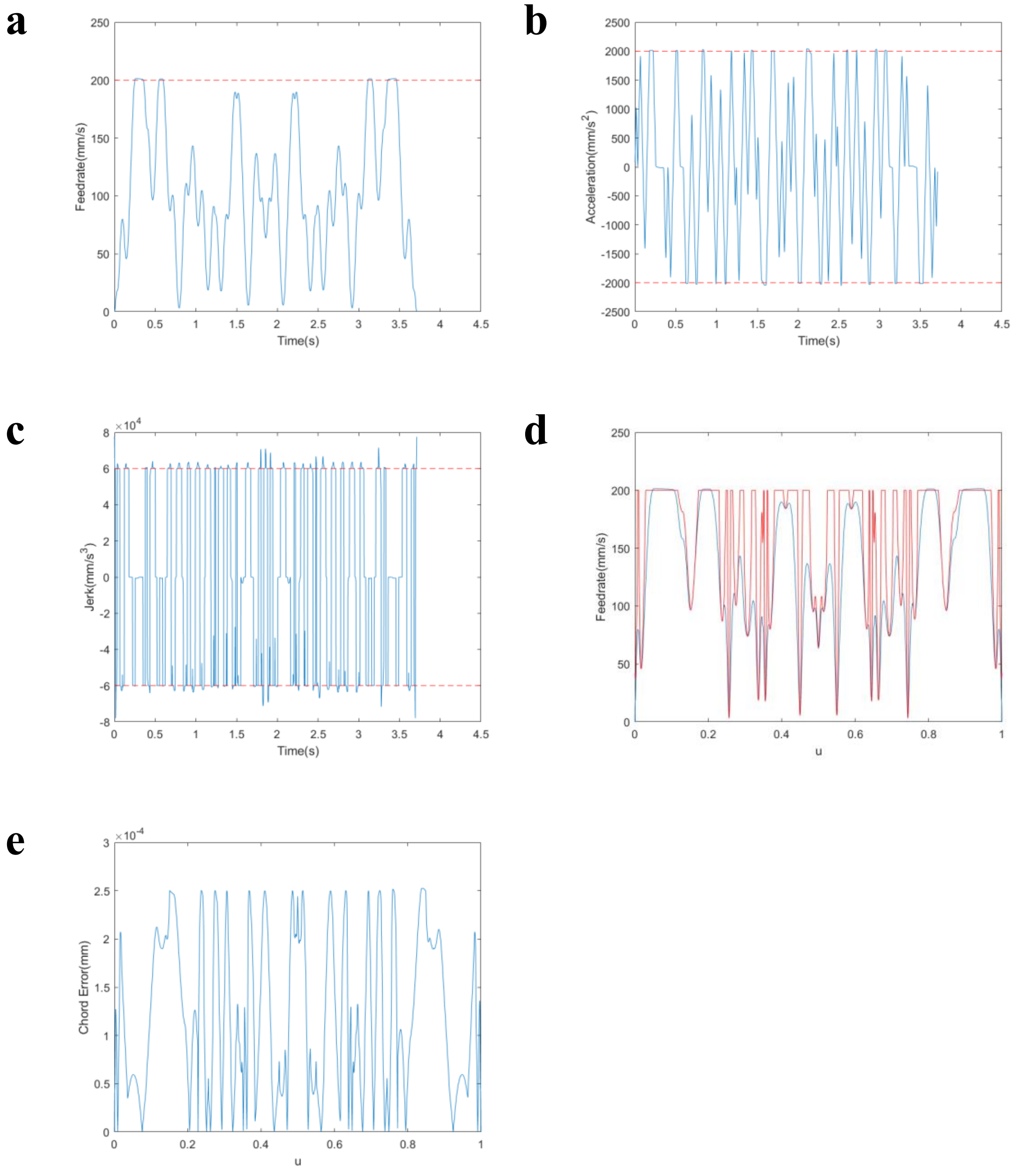

**Fig. 9** Simulation results of butterfly-shaped curve by method in [36]. **a** Scheduled feedrate. **b** Acceleration. **c** Jerk. **d** Comparison of actual feedrate and maximum

allowable feedrate. **e** Chord error.

**Table 2** Static comparison of butterfly-shaped curve simulation results

| Test methods | Maximum federate (mm/s) | Maximum acceleration (mm/s$^2$) | Maximum jerk (mm/s$^3$) | Maximum chord error (mm) | Average chord error (mm) | Interpolation time (s) |
|---|---|---|---|---|---|---|
| **Proposed method** | 200.0000 (↓0.50%) | 2.0000e+3 (↓2.63%) | 5.9973e+4 (↓23.04%) | 2.5000e-4 (↓1.03%) | 9.8172e-5 (↓13.53%) | 4.030 (↑8.03%) |
| **Method in [25]** | 200.0000 | 2.0001e+3 | 6.0139e+4 | 2.5190e-4 | 1.1301e-4 | 3.723 |
| **Method in [36]** | 201.3864 | 2.0504e+3 | 7.7931e+4 | 2.5261e-4 | 1.1353e-4 | 3.712 |

4.2 Analysis and comparisons of trident curve

As shown in Fig.6 b, there are 6 break points on the NURBS curve, and the trident curve is divided into 6 segments. The simulation results obtained through the proposed feedrate scheduling method are shown in Fig.10. From Fig.10 a-c, we know that the feedrate, acceleration and jerk values of proposed interpolation algorithm are controlled within the maximum values. Meanwhile, the acceleration and jerk profile is continuous, which means the feedrate profile is smooth enough. From Fig.10 d, we can notice that the actual feedrate always does not excess the maximum allowable feedrate. As shown in Fig.10 e, the chord errors of all interpolation points are far less than the given threshold. This shows the high accuracy of the proposed method.

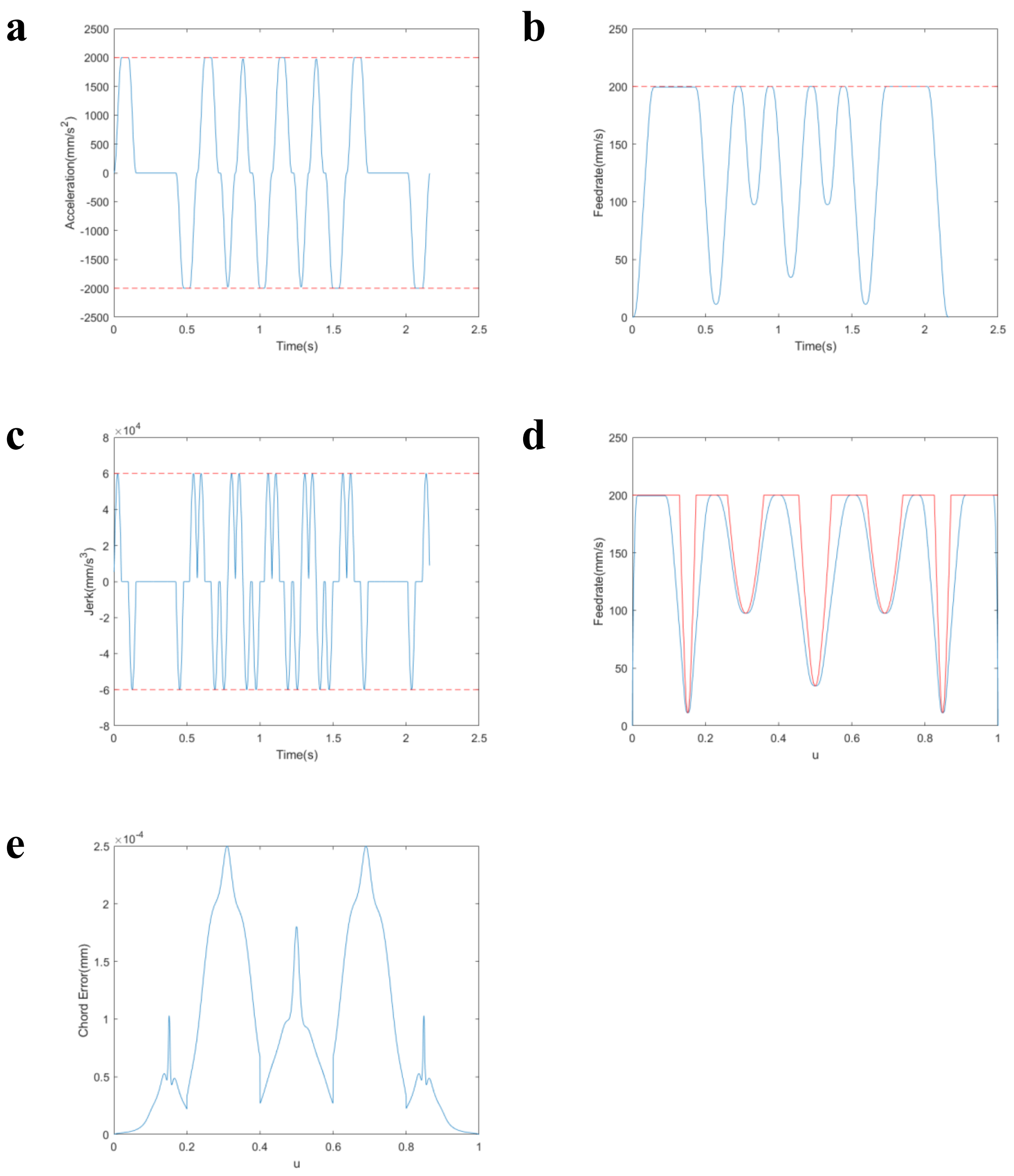

**Fig. 10** Simulation results of trident curve by proposed method. **a** Scheduled feedrate. **b** Acceleration. **c** Jerk. **d** Comparison of actual feedrate and maximum allowable feedrate. **e** Chord error.

Since all segments of the trident curve are long enough to have a constant feedrate section, the method in [32] can been proposed for comparison. As can be seen in Fig 11. b-c, the acceleration and jerk profile are continuous, and the actual maximum acceleration and jerk can reach the maximum acceleration and jerk values in Table 1. However, the actual maximum acceleration is not maintained for a while, which means the kinematic performance of the machine tool having not yet been fully utilized and leads to the increase of interpolating time. This can also be confirmed from Table 3, the total interpolation time of the proposed method is shorter than the method in [32] by 5.71%.

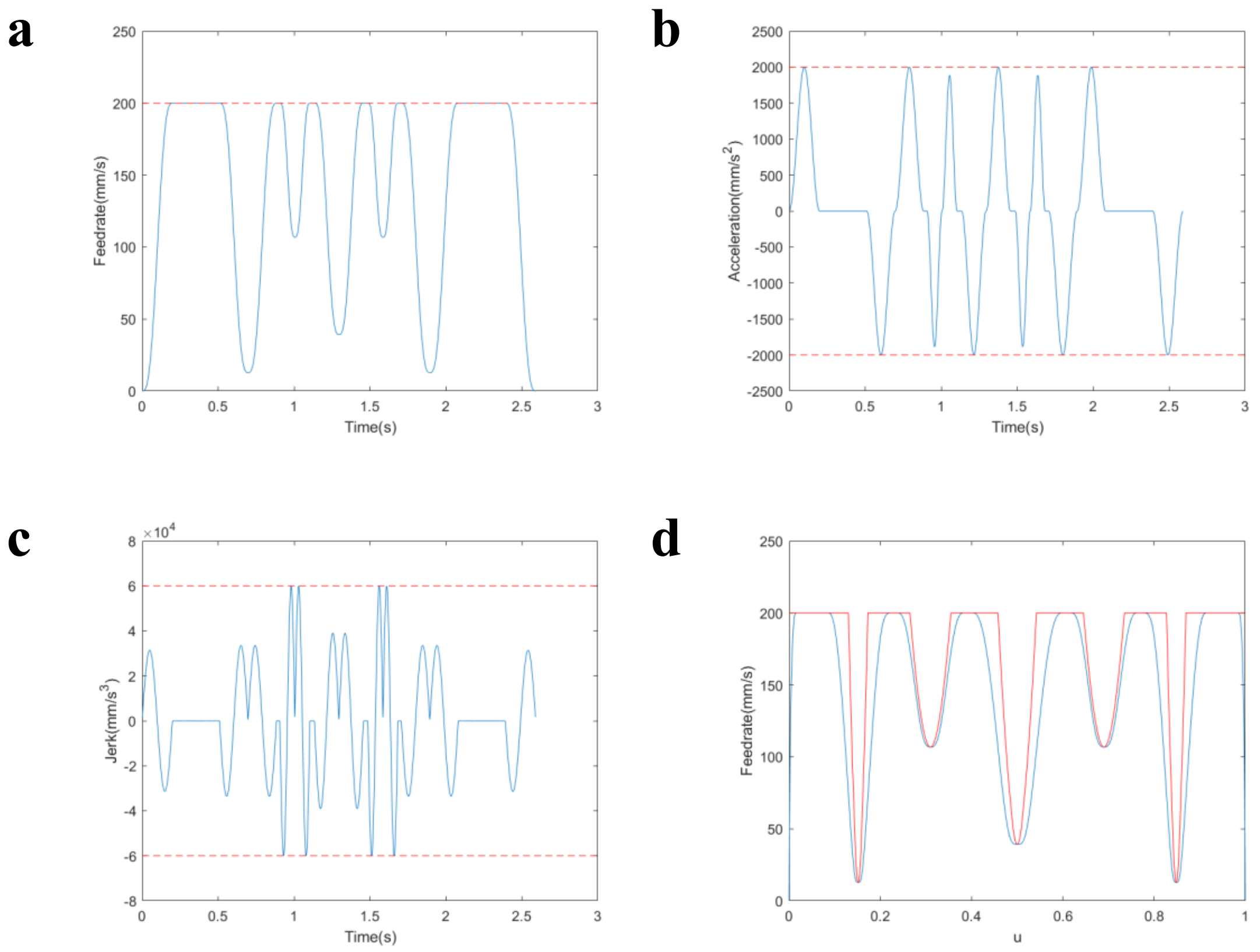

**e**

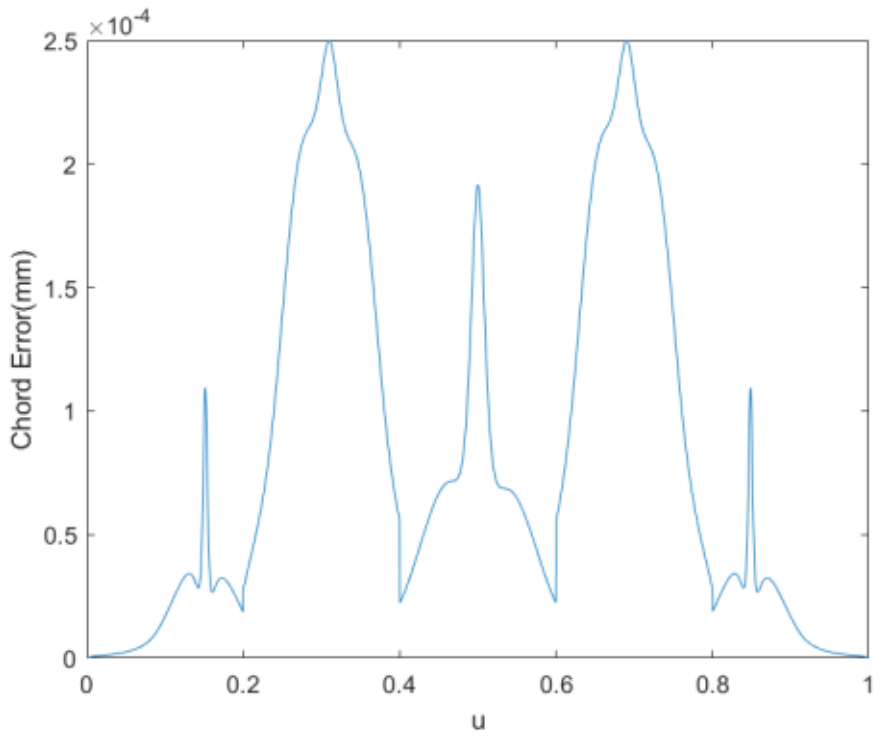

**Fig. 11** Simulation results of trident curve by method in [32]. **a** Scheduled feedrate. **b** Acceleration. **c** Jerk. **d** Comparison of actual feedrate and maximum allowable feedrate. **e** Chord error.

**Table 3** Static comparison of trident curve simulation results

| **Test methods** | **Maximum federate** (mm/s) | **Maximum acceleration** (mm/$s^2$) | **Maximum jerk** (mm/$s^3$) | **Maximum chord error** (mm) | **Average chord error** (mm) | **Interpolation time** (s) |
|---|---|---|---|---|---|---|
| **Proposed method** | 200.0000 | 2.0000e+3 | 5.9973e+4 | 2.4998e-4 | 5.5656e-5 | 2.443 (↓5.71%) |
| **Method in [32]** | 200.0000 | 1.9999e+3 | 5.9970e+4 | 2.4999e-4 | 5.0657e-5 | 2.591 |

## 5. Conclusions and future works

This paper proposed a novel and complete S-shape feed rate scheduling approach which consists of three modules. Firstly, the bidirectional scanning module based on our improved velocity scheduling function is implemented to guarantee the continuity of the feed rate at the junctions between successive NURBS sub-curves. Then, the NURBS sub-curves are classified into two cases (i.e., ACC-or-DEC case and ACC-and-DEC case), which makes the velocity scheduling module easy to implement. Finally, a round-off error elimination module is proposed to decrease the actual maximum feed rate of a certain NURBS sub-curve appropriately, which leads to the total interpolating time becoming an integer multiple of the interpolation period, then the round-off error is eliminated. The benchmarks prove the applicability of the proposed method.

Advantages of our method are summarized as follows:

1. The proposed method is complete, and different velocity scheduling schemes are used for sub-curves of different lengths, which means it can handle more complex curves (such as a butterfly-shaped curve).
2. The proposed method generates a smoother feed rate profile, continuous acceleration curve, and continuous jerk curve. Meanwhile, the actual maximum acceleration and jerk are under the confined range strictly.
3. The total interpolating time is an integer multiple of the interpolation period, which eliminates the round-off error.
4. Compared with the traditional method that can generate a smooth enough feedrate profile, the total interpolating time of the proposed method is shorter.

In future studies, the authors intend to expand the method from the three-axis to the five-axis stage to further develop its potential.

**Acknowledgement**

This work has been supported by the National Key Research and Development Program of China (Grant No.2018YFB1107402), Beijing Natural Science Foundation (Z180005) and the National Natural Science Foundation of China (Grant No.11290141).

**Compliance with Ethical Standards**

The authors declare that they have no conflict of interest that could have direct or potential influence or impart bias on the research reported in this paper. The research does not involve human participants and/or animals. Consent to submit the paper for publication has been received explicitly from all co-authors.

**Appendix**

1. Parameters of butterfly-shaped curve

The degree: p=3.

The control point (mm): P=[(54.493, 52.139), (55.507,52.139), (56.082, 49.615), (56.780, 44.971), (69.575,51.358), (77.786, 58.573), (90.526, 67.081), (105.973,63.801), (100.400, 47.326), (94.567, 39.913), (92.369,30.485), (83.440,

33.757), (91.892, 28.509), (89.444,20.393), (83.218, 15.446), (87.621, 4.830), (80.945, 9.267),(79.834, 14.535), (76.074, 8.522), (70.183, 12.550), (64.171,16.865), (59.993, 22.122), (55.680, 36.359), (56.925, 24.995),(59.765, 19.828), (54.493, 14.940), (49.220, 19.828), (52.060,24.994), (53.305, 36.359), (48.992, 22.122), (44.814, 16.865),(38.802, 12.551), (32.911, 8.521), (29.152, 14.535), (28.040,9.267), (21.364, 4.830), (25.768, 15.447), (19.539, 20.391),(17.097, 28.512), (25.537, 33.750), (16.602, 30.496), (14.199,39.803), (8.668, 47.408), (3.000, 63.794), (18.465, 67.084),(31.197, 58.572), (39.411, 51.358), (52.204, 44.971), (52.904,49.614), (53.478, 52.139), (54.492, 52.139)].

The knot vector: U= [0, 0, 0, 0, 0.0083, 0.015, 0.0361, 0.0855, 0.1293, 0.1509, 0.1931, 0.2273, 0.2435, 0.2561, 0.2692, 0.2889, 0.3170, 0.3316, 0.3482, 0.3553, 0.3649, 0.3837, 0.4005, 0.4269, 0.4510, 0.4660, 0.4891, 0.5000, 0.5109, 0.5340, 0.5489, 0.5731, 0.5994, 0.6163, 0.6351, 0.6447, 0.6518, 0.6683, 0.6830, 0.7111, 0.7307, 0.7439, 0.7565, 0.7729, 0.8069, 0.8491, 0.8707, 0.9145, 0.9639, 0.9850, 0.9917, 1.0, 1.0, 1.0, 1.0].

The weight vector: W= [1.0, 1.0, 1.0, 1.2, 1.0, 1.0, 1.0, 1.0, 1.0, 1.0, 1, 2, 1.0, 1.0, 5.0, 3.0, 1.0, 1.1, 1.0, 1.0, 1.0, 1.0, 1.0, 1.0, 1.0, 1.0, 1.0, 1.0, 1.0, 1.0, 1.0, 1.0, 1.0, 1.1, 1.0, 3.0, 5.0, 1.0, 1.0, 2.0, 1.0, 1.0, 1.0, 1.0, 1.0, 1.0, 1.0, 1.2, 1.0, 1.0, 1.0].

2. Parameters of trident curve

The degree: p=2.

The control point (mm): P={(60, 0), (120, 120), (72, 48), (60, 120), (48, 48), (0, 120), (60, 0)}.

The knot vector: U=[0, 0, 0, 0.2, 0.4, 0.6, 0.8, 1, 1, 1].

The weight vector: W=[1, 1, 1, 1, 1, 1, 1].